\documentstyle	[12pt]{article}
	\let\a=\alpha	\let\b=\beta	\let\g=\gamma
\let\d=\delta	\let\e=\epsilon 
	\let\q=\theta	
 \let\l=\lambda	\let\m=\mu
\let\n=\nu	\let\x=\xi		
\let\s=\sigma		
\let\f=\phi	\let\c=\chi	\let\y=\psi
\let\w=\omega	\let\G=\Gamma	\let\D=\Delta
	\let\L=\Lambda 
\let\P=\Pi	\let\S=\Sigma	
\let\F=\Phi	\let\Y=\Psi 
\let\W=\Omega \let\la=\label	\let\ci=\cite

\let\ti=\tilde

\def\bd{\begin{document}}
\def\ed{\end{document}}
\def\ds{\documentstyle}	\let\fr=\frac
\let\bl=\bigl	\let\br=\bigr \let\Br=\Bigr
\let\Bl=\Bigl \let\bm=\bibitem
\let\na=\nabla \let\pa=\partial
\let\ov=\overline
\newcommand{\be}{\begin{equation}}
\newcommand{\ee}{\end{equation}}
\def\ba{\begin{array}} \def\ea{\end{array}}
\newcommand{\ho}[1]{$\,	^{#1}$}
\newcommand{\hoch}[1]{$\,	^{#1}$}
\newcommand{\bea}{\begin{eqnarray}}
\newcommand{\eea}{\end{eqnarray}}
\newcommand{\ra}{\rightarrow}
\newcommand{\lra}{\longrightarrow}
\newcommand{\Lra}{\Leftrightarrow}
\newcommand{\ap}{\alpha^\prime}
\newcommand{\bp}{\beta^\prime}
\newcommand{\tr}{{\rm	tr}	}
\newcommand{\Tr}{{\rm	Tr}	}
\newcommand{\NP}{Nucl.	Phys.	}
\newcommand{\tamphys}{\it	Center	for
Theoretical	Physics\\ Physics	Department	\\
Texas	A	\&	M	University \\	College	Station,
Texas	77843}

\begin{document}

\hfill{CTP-TAMU-46/93}

\hfill{hep-ph/9309254}

\vspace{24pt}

\begin{center} { \large {\bf On-shell Supersymmetry Anomalies
and the Spontaneous Breaking of Gauge Symmetry }}

\vspace{36pt}

J. A. Dixon\footnote{Supported in part by the U.S. Dept of
Energy, under grant DE-FG05-91ER40633, Email: dixon@phys.tamu.edu}

\vspace{6pt}

{\tamphys}

\vspace{6pt}

September, 1993

\vspace{6pt}

\underline{ABSTRACT}

\end{center}

A search for supersymmetry anomalies requires an
examination of the BRS cohomology of supersymmetric Yang-Mills coupled
to chiral matter, and the physically interesting
(on-shell) anomalies are those which cannot be eliminated using the
equations of motion. An analysis of this cohomology problem shows
that the simplest situation where a physically interesting
supersymmetry anomaly can arise  is when:
\begin{enumerate}
\item
the anomaly occurs in the renormalization of a
composite antichiral spinor superfield operator
constructed from the chiral matter in the theory,
\item
the anomalous diagrams are one-loop triangle diagrams containing
a gauge propagator,
\item
the gauge symmetry (but not supersymmetry)
is spontaneously broken,
\item  the initial operator has a dimension such
that the triangle diagram is at least linearly divergent,
\item
the anomaly contains only massless chiral superfields,
although (apparently harder to calculate) supersymmetry
anomalies can also contain massive chiral superfields,
 \item
the theory contains vertices which,
after gauge symmetry breaking,
couple massless matter fields to
massive matter fields and massive gauge fields, like the
$ e^- \n_e W^+$ vertices of the standard model.
\end{enumerate}

The supersymmetry anomalies considered here are all
`soft', in the sense that they must vanish when certain masses go
to zero. It appears that the order parameter of the resulting
supersymmetry breaking may be the vacuum expectation value that
breaks the gauge symmetry.
Nonzero values for the anomalies, if they exist,
appear to generate
supersymmetry breaking for observable particles with
a cosmological constant that is naturally zero.

A specific example of a possibly anomalous
operator which arises in this way is then
examined.   An analysis is made of the most efficient way
to try to calculate supersymmetry anomalies for a simple
case. It remains to do some careful analyses and Feynman graph
calculations of the relevant  coefficients
for some specific examples.

\vfill

\baselineskip=24pt

\pagebreak

\tableofcontents

\setcounter{page}{1}

\section{Introduction and Discussion}

Supersymmetry breaking is considered by many physicists to be
one of the chief problems in modern  elementary particle theory.  It
is amusing and  ironical that this is so, since there is so little
phenomenological  evidence for supersymmetry.
The reason for this strange situation is of course that
supersymmetry is theoretically very attractive. Some of the
problems of dynamical supersymmetry breaking and some
references to the recent literature can be found in
\ci{banks}.  See also \ci{brignole} for a recent
discussion of the `soft terms' approach to supersymmetry
breaking.

Recently it was suggested \ci{erice} that there might
be a mechanism whereby supersymmetry breaks itself through
supersymmetry anomalies.  This would be a pretty phenomenon if
it works, because it would be calculable in
perturbation theory, inevitable rather than contrived,
and it also leaves the cosmological constant
at a zero value in a natural way after supersymmetry breaking,
assuming it was zero before supersymmetry breaking, which is
often the case in superstring theory.  This  mechanism
for supersymmetry breaking affects only composite
operators which can act as interpolating fields for the
bound states (like the hadrons) or the broken phase
states (like the electron), because this is where the
supersymmetry anomalies can occur, according to the
BRS cohomology results.

These superanomalies can occur only in certain composite
operators in supersymmetric theories.  This explains why
they have escaped detection up to now, even if they
actually exist with non-zero coefficients.  It is
a real chore to find them.  This chore requires a lot of
knowledge of the BRS cohomology space before one even
begins a Feynman graph calculation.  Then the calculation
is very model dependent and also depends on the operator
chosen.  Also, since the effect is mass dependent, the
number of terms and the
integrals involved appear to make the problem fairly labour intensive.
For the example given below in section \ref{secexample},
the individual diagrams that
need calculation are certainly not zero, but they might nevertheless
give a zero anomaly after regularization, addition and variation
by $\d$. I do not at present know a reliable method
to try to calculate supersymmetry anomalies or alternatively show that they
are absent. It does appear to
be impossible to find a regularization procedure that respects both
gauge invariance and supersymmetry, so that it seems possible
that the anomalies discussed here do appear with non-zero
coefficients. It is tantalizing that the superanomalies do seem to
require both gauge invariance and supersymmetry for purely
cohomological reasons as explained below.
However, it may be that the coefficients are always
zero.  How could the regularization problem give rise
to mass dependent anomalies?  Is there some sort of choice
between gauge anomalies and supersymmetry anomalies going on
here?  These seem to be hard questions.
The first step at any rate is to do the present cohomology analysis
in order to even know which diagrams to calculate.  The problem of
calculating them must come later.

So the purpose of this paper is to analyze the BRS
cohomology to the point necessary to have a
reasonable chance of finding a superanomaly.
Some of the discussion of this section and the
bare bones of the example to follow formed part
of a recent talk \ci{tamu45}. The present paper contains
more complete results.

To motivate the problem analyzed in this paper, I shall try to
explain some
of the fairly obvious reasons that such
anomalies are not present in some
simple cases. Then we will examine a
case which appears to require a detailed
calculation to determine whether a
supersymmetry anomaly is or is not present.

It is easy to write down the simplest
examples where a supersymmetry anomaly could
conceivably arise. The BRS cohomology of
these theories indicates that there could be
anomalies in the renormalization of
composite operators (made from the
elementary chiral superfields $S$ of the
theory) which are antichiral spinor
superfields.  These composite operators satisfy the antichiral constraint:
\be
D_{\a} \Y_{\b} =0
\ee
and take forms such as:
\be
\Y_{1 \a} = D^2 [ S_1 D_{\a} S_2 ]
\la{e1}
\ee \be
\Y_{2 \a} = {\ov S}_1 D^2 [ S_2  D_{\a} S_3 ]
\ee
\be  \Y_{3 \a} = {\ov S}_1 D^2 [ {\ov
D}^2 {\ov S}_1   D_{\a} S_3 ]
\la{e3}
\ee
\be  \Y_{4 \a} = D^2 \{
{\ov D}^{\dot \b} {\ov S}_1  {\ov
D}_{\dot \b} D_{\a} S_2 \}
\la{e4}
\ee
One could
add more chiral superfields $S$ or more
supercovariant derivatives of course.  The
main things to  keep in mind are:
\begin{enumerate} \item The expression for
$\Y_{ \a}$ should not vanish
\item It is frequently necessary to use more
than one flavour of superfield $S$ to
prevent the expression from vanishing,
because such expressions may be
antisymmetric under interchange of flavour
indices
\item It is probably necessary that the resulting integral should
be at least linearly divergent to give rise to an anomaly,
though it is not entirely clear that this is either necessary or
sufficient
\end{enumerate}

To calculate the anomaly, one would couple
such composite operators to the action with
an elementary (i.e. not composite)
antichiral spinor source superfield  $\F^{\a}$.
This means that one simply adds the following
term to the usual action of the theory:
\be
S_{\rm \F}  = \int d^6 {\ov z} \; \F^{\a}
\Y_{\a} \la{action}
\ee
where $\Y_{\a}$ is some composite antichiral
spinor superfield, some examples of which are given above
in (\ref{e1}-\ref{e4}).

Then the anomaly would appear in the form:
\be  \d \G_{\F} = m^k \int d^6 {\ov z} \; \F^{\a}
c_{\a} {\ov S}^n   \la{anom} \ee  where $\G_{\F}$
is the one-particle irreducible generating
functional with one insertion of the source
$\F_{\a}$, $\d$ is the nilpotent BRS
operator, $\int d^6 {\ov z}$ is an integral
over antichiral superspace, $c_{\a}$ is the
constant ghost parameter of rigid
supersymmetry, $m^k$ is the  mass parameter
$m$ to some power $k$ required by simple
dimensional analysis, and  ${\ov S}^n $ is
the  $n^{th}$ power of the antichiral
superfield ( this might include  a sum over
indices which distinguish different $\ov S$
superfields from each other).

To count masses we use the following
assignments for the variables and the
superfields:
\be
  m=1;\pa_{\m} =1; \q_{\a}=\fr{-1}{2}; S=1;
c_{\a}=\fr{-1}{2};
\F_{\a}= \fr{1}{2}; \ee
Now we define the component fields:
\be
S(x, \q, {\ov \q}) = A(y) + \q^{\a} \y_{\a}(y) + \fr{1}{2} \q^2 F(y)
\ee
where
\be
y^{\m} = s^{\m} + \fr{1}{2} \q^{\a} \s^{\m}_{\a \dot \b}
{\ov \q}^{\dot \b}
\ee
satisfies
\be
{\ov D}_{\dot \a} y^{\m} =
{D}_{\a} {\ov y}^{\m} = 0
\ee
Similarly we have:
\be
\F_{\a}(x, \q, {\ov \q})
= \f_{\a}({\ov y}) + W_{\a \dot \b}({\ov y}) {\ov \q}^{\dot \b}
+ \fr{1}{2} {\ov \q}^2 \c_{\a}({\ov y})
\ee
The dimensions of these component fields are then:
\be
  A=1; \y_{\a} =
\fr{3}{2}; F=2;   \c_{\a}=
\fr{3}{2};\f_{\a}= \fr{1}{2};W_{\a \dot \b}=
1.
\ee

An examination of  examples shows that
elementary dimensional  counting prevents
the powers of $m$ from working correctly to
yield (\ref{anom})
whenever the only vertices
of the diagram are chiral vertices involving
only chiral fields.  It should be possible to show
this by a dimensional argument, but this has
not yet been done--at any rate it certainly
seems to hold for a wealth of examples, one of which can
be found in \ci{erice}.

  However when there is at
least one gauge propagator in the diagram,
the powers of $m$ easily work out correctly to
yield (\ref{anom}).   But then one has to
confront another problem, which is that one
has to analyze the cohomology again in the
presence of the gauge fields.
This unsolved problem has been partially and sufficiently finessed in
the present paper.

Another problem that was also unsolved and also necessary for
our present purposes is the
problem of solving the full BRS cohomology of
any supersymmetric theory including the
sources that are necessary to formulate the
full BRS identity.  Essentially, this brings
in the complication of ensuring that the BRS
cohomology space is orthogonal to the equations
of motion of the fields.  This is the main
subject of the present paper.

Whenever one formulates a BRS identity in
the manner pioneered by Zinn-Justin, it is necessary to
also include sources ${\tilde f}_i$ for the
variation of the fields $f_i$, and in the
resulting `full' BRS operator, these give
rise to terms that involve the equations of
motion of the corresponding fields.    This
turns out to be  more or less equivalent to
the Batalin-Vilkovisky quantization method.
The essential point is that this eliminates
from the cohomology space anything which
vanishes by the equation of motion, i.e.
anything which vanishes `on-shell'.
At the simplest level, as explained more fully
below,  this will eliminate
all those objects in the cohomology space
which involve superfields $\ov S$ which have
mass terms in the action, as well as a
number of higher order terms that are of no
concern at present.

So we are now interested only in computing
diagrams where the possible supersymmetry
anomaly involves massless antichiral fields $\ov
S$ in (\ref{anom}).  But this raises another
problem.  The  most promising simple  case
(see below) seems to involve   a triangle diagram with
the $\F^{\a}$ superfield at one vertex, two
chiral (or antichiral) superfields emerging
from that vertex and the exchange of a
vector superfield between these two lines.
Now the mass counting implies that the
anomaly (\ref{anom}) generally has a higher power of mass than the
composite operator (\ref{action}) from which it arises.
The only way this can happen is if some of
the interior lines are massive.  Is there
any  way that interior massive lines can
give rise to exterior massless lines while
exchanging a vector superfield? The answer to
this question is of course well known--this
will happen if and only if  the gauge
symmetry is spontaneously broken, provided the representations
are chosen correctly, as is discussed below.  We will therefore
assume that gauge symmetry {\em is} spontaneously broken and that
supersymmetry {\em is not}
spontaneously broken. Since we are looking for supersymmetry
breaking through anomalies, it is reasonable to assume that
it is not otherwise broken.

 This combination is
in fact very easy to achieve--as is well
known \ci{oraif}, gauge symmetry breaking is natural
and very easy to achieve  in rigid
supersymmetry, but spontaneous supersymmetry
breaking can only be achieved with very
contrived models, particularly if the gauge
group is semisimple.

So now, if we want to examine the question
of supersymmetry anomalies, we are forced to
consider a supersymmetric gauge theory with
spontaneous breaking of the gauge symmetry.
But there are more conditions, at least for
the supersymmetry anomalies that involve massless
matter superfields.  In order for the
relevant diagrams to exist, we must have
matter multiplets which break under the gauge
breaking into a combination of massive and
massless fields, so that a massive vector superfield
can have a vertex with a massless and a massive chiral
superfield.

This happens of course for
the Higgs multiplet itself, but then the
massless Goldstone supermultiplets do not
contribute to the relevant BRS cohomology space, as
will be shown below.
We must have additional
(non-Higgs) matter multiplets which break
under the gauge breaking into a combination
of massive and massless fields.  There are
many ways to do this, and an example is
given below.  Note that this happens also in
the standard model, where  the neutrino
remains massless after spontaneous breaking
of $SU(2) \times U(1)$ to $  U(1)_{EM}$
simply  because there is no right handed
neutrino for it to  form a mass with (and
because lepton conservation prevents the
formation of a Majorana neutrino mass, in
the minimal standard model at least). The
relevant discussion of the
standard model will be the subject of a forthcoming
paper \ci{tamu47}.

So if we want to find a simple supersymmetry anomaly,
we are driven to models with gauged
supersymmetry and spontaneous breaking of
the gauge symmetry through Higgs multiplets
which develop a VEV in their `A' components
(but not their `F' components--that would
break supersymmetry).  In addition these
models must have matter which is massless at
tree level, but which gets split into
massive and massless components as a result
of gauge breaking.    These are the simplest
 models that have a chance of developing
supersymmetry anomalies in some of their
composite operators at the one loop level.  Such models are of
course  very  reminiscent of a
supersymmetric version of the standard model
of strong, weak and electromagnetic
interactions.  It is just within the realm
of possibility that these anomalies could
account for the experimentally observed lack
of  supersymmetry  in the world with no
additional  assumptions in the model at
all--in which case we could say that
supersymmetry breaks itself.  But there is
plenty of work to do before we can determine
whether this notion is right.  Even if the supersymmetry anomalies
exist, considerable work will be necessary to deduce the
form of the
supersymmetry breaking they give rise to.

A rather interesting and new feature is that
we can see that the particular `soft'
mass-dependent supersymmetry anomalies we
are examining here, if their coefficients are non-zero, would
give rise to a kind of supersymmetry
breaking that is a function of the VEV
that    breaks the  gauge symmetries, and
which vanishes in the gauge symmetric
limit.     This is still consistent with the
conjecture \ci{erice} mentioned above that such anomalies might also provide a
natural mechanism whereby `supersymmetry breaks itself', while at
the same time retaining the cosmological constant at the zero value
it naturally has in many unbroken supersymmetric theories.
Spontaneous breaking of the gauge symmetry would
not interfere with this feature, because it does not change
the vacuum energy so long as supersymmetry is not spontaneously broken
at the same time.

   There is still a possibility of
`hard' supersymmetry anomalies too, which we
do not consider here, since they look more
difficult to compute.

\section{Supersymmetric gauge theory with
spontaneous breaking of gauge symmetry}

Pursuant to the above discussion,  we will
now consider a general  supersymmetric gauge
theory coupled to chiral matter, where the
gauge symmetry is spontaneously broken.  We
will consider this action in terms of
component fields in the Wess-Zumino gauge,
and we will fix the gauge in the way
pioneered by `t Hooft for spontaneously
broken theories.  Naturally much of this is
simpler in superspace, but the method we use
to find cohomology actually brings us back
to components anyway, and also it is
probably better to use components to compute
diagrams when one is looking for something
as obscure and tricky as anomalies.

The action consists of the following parts,
each of which is separately supersymmetric
and gauge invariant (before spontaneous
breaking): \[ S_{\rm	Total}	= S_{\rm	YM	} +
S_{\rm	Matter} + S_{\rm	Chiral}
\] \be + {\ov	S}_{\rm	Chiral}
+ S_{\rm	Ghost} + S_{\rm	Sources}
+ S_{\F }
+ {\ov	S}_{\ov \F}
\la{stot}
\ee We
will discuss each of these terms in the next
section.

\section{Component Form of Action}

The Yang-Mills supersymmetric action in the
Wess-Zumino gauge takes the form:
 \be S_{\rm	YM}	=	\int	d^{4}x	\Bigl	\{
-\frac{1}{4}	G_{\mu	\nu}^{a}	G^{a	\mu	\nu} -
\frac{1}{2}	\lambda^{a	\a} \s^{\m}_{\a
\dot{\b}	} D_{\m}^{a	b}	\ov{\lambda}^{b
\dot{\b}	} +	\frac{1}{2}	D^{a}D^{a} \Bigr	\}
\la{actym}
\ee where we will assume that the gauge group
is semisimple, and
\be G^a_{\m\n} = \pa_{\m}
V^a_{\n}  - \pa_{\n} V^a_{\m} + f^{abc}
V^b_{\m} V^c_{\n} \ee \be D_{\m}^{a	b}
\ov{\lambda}^{b	\dot{\b}	} =  \pa_{\m}
\ov{\lambda}^{a	\dot{\b}	} +  f^{abc}
V^b_{\m}  \ov{\lambda}^{c	\dot{\b}	} \ee

We will assume that the matter is in some
(generally reducible) representation of the
compact semi-simple gauge group.  The
matrices $T$ satisfy: \be [ T^{a }, T^{b}] =
i f^{abc} T^c  \ee and have indices of the
form \be T^{ai}_{\;\;\; j} \ee In general
these matrices are Hermitian complex
matrices, which means that:
\be
[T^{ai}_{\;\;\; \;\;j}]^*
=
{\ov T}^{a\;\;\;\;\;j}_{\;\;\; i}
=
T^{aj}_{\;\;\;\;\; i}
\ee
We will assume that the VEV of the
`A' component takes the form \be <A^i> = m u^i
\ee where $u^i$ is a dimensionless quantity
that describes the direction and magnitude
in  group space of the breaking, and $m$ is
the mass parameter of the theory.  Then the
`kinetic' part of the matter action is, after a shift:
 \[
S_{\rm	Matter}	=	-	\int	d^{4}x	\Bigl	\{
D^i_{\mu	j}	(m	u^j	+	A^j)	{\ov	D}_i^{\m	k	}
(m	{\ov	u}_k	+	{\ov	A}_k) \] \[ +	\y^{\a	i}
\sigma^{\m}_{\;\a	\dot	\b}	(D_{\mu}
\ov{\y}^{\dot	\b})_i -	F^i	\ov{	F}_i +
\y^{\a	j}	T^{ai}_{j}	\l^a_{\a} (m	{\ov	u}_i
+	{\ov	A}_i) \] \[ +	{\ov	\y}^{\dot	\a}_{	j}
T^{aj}_{i}	{\ov	\l}^a_{\dot	\a	} (m	u^i	+	A^i)
\] \be + D^a	T^{aj}_{i} (m	{\ov	u}_j	+	{\ov
A}_j) (m	u^i	+	A^i) \Bigr	\}
\la{actmat}
\ee
Next we must
discuss the chiral part of the action. We assume that
it is such that there is a minimum at some
non-zero VEV as discussed above.  We will
consider non-renormalizable terms up to fourth order here
just to show how they would work, though
a renormalizable theory would probably be sufficient for all
purposes.

After a
shift, this takes the form \[ S_{\rm	Chiral}
=	\int	d^{4}x	\Bigl	\{ m	g_{ij}(u)[2 A^i	F^j	+
\y^{\a	i}	\y_{\a}^j]
 \] \be +
3 g_{ijk}(u)	[A^i	A^j	F^k	+	A^i	\y^{\a	j}
\y_{\a}^k ]
+\fr{1}{m}  g_{ijkl}	[4 A^i	A^j	A^k  F^l	+	6 A^i A^j
	\y^{\a	k}
\y_{\a}^l	] \Bigr	\}
\la{actchiral}
\ee where
\be g_{ij}(u)
= g_{ij} + 3 g_{ijk} u^k
+ 6 g_{ijkl} u^k u^l \ee
\be g_{ijk}(u)
= g_{ijk}
+ 4 g_{ijkl} u^l \ee
Here $g_{ij}$,
$g_{ijk}$ and $g_{ijkl}$ are dimensionless group invariant tensors under
the action of the matrices $T$. The
parameter $u^i$ in the above is a solution
of the equations:
\be < {\ov F}_i> = 2 g_{ij} u^j
+ 3 g_{ijk}  u^j u^k
+ 4 g_{ijkl} u^j u^k
u^l =0
\ee
\be
< D^a> = T^{ai}_{\;\;j} {\ov u}_i u^j =0
\ee
These equations state that the VEV of the auxiliary fields $F$ and
$D$ are zero. This in turn insures
 that supersymmetry is preserved even though
there is a non-zero VEV $<A^i> = m u^i$ of the unshifted
scalar field $A^i$.
We assume here for simplicity that there are no linear terms in the chiral
action.

Now let us consider the ghost and gauge-fixing action:
\[
S_{\rm	Ghost}	=	-	\delta \int	d^{4}x
	\Bigl	\{	\xi^{a}	[ \frac{k}{2}	Z^{a}	+
\partial^{\mu}	V_{\mu}^{a} +	k	\;	m	\;
T^{aj}_i	u^i	{\ov	A}_j +	k	\;	m	\;	{\ov
T}^{aj}_i	{\ov	u}_j	{A}^i	] \Bigl	\} \] The
field $\w^a$ (used below) is the (real scalar) Fadeev-Popov ghost and
$\x^a$ is the  (real scalar) Fadeev-Popov antighost.
$Z^a$ is a (real scalar)  auxiliary field used for
convenience in gauge fixing.  The $u^i$
dependent terms are included to eliminate
mixing between gauge bosons and Goldstone
bosons in the gauge fixed theory \ci{thooft}.  k is a
gauge parameter, and any physically
meaningful quantity (such as the coefficient
of an on-shell supersymmetry anomaly) should be
independent of it.   Here $\d$ is the BRS
variation of the relevant terms, given
below. When this is expanded out using the
formulae below in section (\ref{full}), we find: \[ S_{\rm	Ghost} =
\int	d^{4}x	\Bigl	\{ (Z^{a}	+	\e^{\m}
\pa_{\m}	\x^a	) [	\frac{k}{2}	Z^{a}	+
\partial^{\mu}	V_{\mu}^{a} +	k	\;	m	\;
T^{aj}_i	u^i	{\ov	A}_j +	k	\;	m	\;	{\ov
T}^{aj}_i	{\ov	u}_j	{A}^i	] \] \[ +	\xi^{a}
\frac{k}{2} (	c^{\a}	\s^{\mu}_{\a	\dot{\b}	}
\ov{c}^{\dot	\b} \pa_{\m}	\x^a	+	\e^{\m}
\pa_{\m}	Z^a	) \] \[ +	\partial^{\mu}	\x^a [
D_{\mu}^{ab}	\omega^{b}	+ \fr{1}{2}	c^{\a}
\s^{\mu}_{\a	\dot{\b}	}	\ov{\lambda}^{a
\dot{\b}	} + \fr{1}{2}	\l^{a	\a	}
\s^{\mu}_{\a	\dot{\b}	} \ov{c}^{\dot{\b}	} +
\epsilon^{\nu}	\partial_{\nu}	V_{\mu}^{a}	]
\] \[ +	k	\;	m	\;	\x^a	T^{aj}_i	u^i [	{\ov
c}^{\dot	\a}	{\ov	\y}_{	j	\a} +	T^{a	k}_j
\w^a	(	m	\,	{\ov	u}_k	+	{\ov	A}_k) +
\epsilon^{\mu}	\partial_{\mu}	{\ov	A}_j] \]
\be +	k	\;	m	\;	\x^a	{\ov	T}^{aj}_i	{\ov	u}_j
[	c^{\a}	\y^i_{\a} +	T^{ai}_j	\w^a	(	m\;	u^j
+	A^j) +	\epsilon^{\mu}	\partial_{\mu}	A^i	]
\Bigl	\} \ee All	the	$\e^{\m}$	terms	in
$S_{\rm	Ghost}$	cancel.  So finally this
takes the form:
 \[ S_{\rm	Ghost} =	\int	d^{4}x	\Bigl	\{ Z^{a}
[	\frac{k}{2}	Z^{a}	+	\partial^{\mu}
V_{\mu}^{a} +	k	\;	m	\;	T^{aj}_i	u^i	{\ov
A}_j +	k	\;	m	\;	{\ov	T}^{aj}_i	{\ov	u}_j
{A}^i	] \] \[ +	\xi^{a}	\frac{k}{2} (	c^{\a}
\s^{\mu}_{\a	\dot{\b}	}	\ov{c}^{\dot	\b}
\pa_{\m}	\x^a	) \] \[ +	\partial^{\mu}	\x^a [
D_{\mu}^{ab}	\omega^{b}	+ \fr{1}{2}	c^{\a}
\s^{\mu}_{\a	\dot{\b}	}	\ov{\lambda}^{a
\dot{\b}	} + \fr{1}{2}	\l^{a	\a	}
\s^{\mu}_{\a	\dot{\b}	} \ov{c}^{\dot{\b}	}	]
\] \[ +	k	\;	m	\;	\x^a	T^{aj}_i	u^i [	{\ov
c}^{\dot	\a}	{\ov	\y}_{	j	\a} +	T^{a	k}_j
\w^a	(	m	\,	{\ov	u}_k	+	{\ov	A}_k)] \] \be +
k	\;	m	\;	\x^a	{\ov	T}^{aj}_i	{\ov	u}_j [
c^{\a}	\y^i_{\a} +	T^{ai}_j	\w^a	(	m\;	u^j	+
A^j)	] \Bigl	\}
\la{actghost}
\ee
Now we come to the
`Source' part of the action.  This is
necessary in order to formulate the Ward
identity for the theory in the form first
advocated by Zinn-Justin \ci{zinn}.  We introduce
sources $\tilde f$ for the BRS variation $\d f$ of each field
$f$ in the action.
The (Bose-Fermi) statistics of the
source is opposite to that of the field, so
that the action has even statistics
(since $\d$ is odd).
\[
S_{\rm	Sources}
	=	\int	d^{4}x \Bigl	\{ {\tilde	F}_i	\d	A^i +
{\tilde	\y}^{\a}_i	\d \y_{\a}^i +	{\tilde
A}_i	\d	F^i \] \[ +	{\tilde	{\ov	F}}^i	\d
{\ov	A}_i +	{\tilde	{\ov	\y}}_{\dot	\a}^i	\d
{\ov	\y}^{\dot	\a}_i +	{\tilde	{\ov	A}}^i	\d
{\ov	F}_i \] \[ +	{\tilde	V}^{	a	\m} \d
V^a_{\m} +	{\tilde	\l}^{a	\a}	\d	\l^a_{\a} +
{\tilde	{\ov	\l}	}^{a	\dot	\a}	\d	{\ov
\l}^a_{\dot	\a} +	{\tilde	D^a	}	\d	D^a \]
\be
+	{\tilde	\w}^a	\d	\w^a +	{\tilde	\x}^a	\d
\x^a +	{\tilde	Z}^a	\d	Z^a \Bigr	\} \ee When
expanded using the form of $\d$ below, this
takes the form: \[ S_{\rm	Sources}
	=	\int	d^{4}x \Bigl	\{ {\tilde	F}^i	[	c^{\a}
\y^i_{\a} +	T^{ai}_j	\w^a	(	m	\;	u^j
+	A^j) +
\epsilon^{\mu}	\partial_{\mu}	A^i] \] \[ +
{\tilde	\y}_i^{\a}	[	(D^j_{\m	k}
(	m	\;	u^k	+
A^k) \s^{\m}_{\a	\dot{\b}}
\overline{c}^{\dot{\b}} +	F^i	c_{\a} +
T^{ai}_j	\w^a	\y_{\a}^j +	\epsilon^{\mu}
\partial_{\mu}	\y_{\a}^i] \] \[ + {\tilde
A}_i	[	(	D_{\m}	\y)^{i	\a} \s^{\m}_{\a
\dot{\b}}	\overline{c}^{\dot{\b}} +	T^{ai}_j
\w^a	F^j +	\epsilon^{\mu}	\pa_{\mu}	F^i] \]
\[
+	{\tilde	{\ov	F}}^i	[	{\ov	c}^{\dot	\a}
	{\ov	\y}^i_{\dot	\a}	 +	{\ov	T}^{aj}_i
\w^a	(	m	\;	{\ov	u}_j	+	{\ov	A}_j) +
\epsilon^{\mu}	\partial_{\mu}	{\ov	A}_i] \]
\[ +	{\tilde	{\ov	\y}}^{i	\dot	\a} [	({\ov D}^j_{\m
i}	(	m	\;	{\ov	u}_j	+	{\ov	A}_j) {\ov
\s}^{\m}_{\dot	\a	\b}
	{c}^{	\b	} +	{\ov	F}_i	{\ov	c}_{\dot	\a} +
{\ov	T}^{aj}_i	\w^a	{\ov	\y}_{j\dot	\a} +
\epsilon^{\mu}	\partial_{\mu}	{\ov	\y}_{i
\dot	\a}] \]
\[ +	{\tilde	{\ov	A}}^i
	[	(	{\ov	D}_{\m}	{\ov	\y})_{i}^{	\dot	\a}
{\ov	\s}^{\m}_{\dot	\a	\b	}	{c}^{	\b} +	{\ov
T}^{a	j}_i	\w^a	{\ov	F}_j +	\epsilon^{\mu}
\pa_{\mu}	{\ov	F}_i	] \] \[ +	{\tilde	V}^{\mu
a}	[	D_{\mu}^{ab}	\omega^{b}	+ \fr{1}{2}
c^{\a}	\s^{\mu}_{\a	\dot{\b}	}
\ov{\lambda}^{a	\dot{\b}	} + \fr{1}{2}	\l^{a
\a	} \s^{\mu}_{\a	\dot{\b}	} \ov{c}^{\dot{\b}
} +	\epsilon^{\nu}	\partial_{\nu}
V_{\mu}^{a}	] \] \[ +	{\tilde	\lambda}^{	a	\a}
[	\frac{1}{2}	G_{\mu\nu}^{a}	\s^{\mu
\nu}_{\a	\b}	c^{\b} -	f^{abc}
\lambda^{b}_{\a}	\omega^{c}	+	i	D^{a}	c_{\a}
+	\epsilon^{\nu}	\partial_{\nu}
\lambda^{a}_{\a}	] \] \[ +	{\tilde	{\ov	\l	}
}^{a	\dot	\a} [	\frac{1}{2}	G_{\mu\nu}^{a}
\ov{\s}^{\mu	\nu}_{\dot{\a}	\dot{\b}	}
\ov{c}^{\dot	\b} -	f^{abc}
\ov{\lambda}^{b}_{\dot	\a}	\omega^{c} -	i
D^{a} {\ov	c}_{\dot{\a}	} +	\epsilon^{\nu}
\partial_{\nu}	{	\ov	\lambda}^{a}_{\dot	\a}	]
\] \[ + {\tilde	D}^{a}	[ \fr{-i}{2}	c^{\a}
\s^{\mu}_{\a	\dot{\b}	} D_{\m}^{ab}
\ov{\l}^{b	\dot{\b}	} +\fr{i}{2}	D_{\m}^{ab}
\l^{b	\a	} \s^{\mu}_{\a	\dot{\b}	}
\ov{c}^{\dot	\b} +	f^{abc}	D^{b}	\omega^{c} +
\epsilon^{\nu}	\partial_{\nu}	D^{a}] \] \[ +
{\tilde	\omega}^{a}	[	-	\frac{1}{2}	f^{abc}
\omega^{b}	\omega^{c} +	c^{\a}	\s^{\mu}_{\a
\dot{\b}	}	\ov{c}^{\dot	\b}	V_{\mu}^{a} +
\epsilon^{\nu}	\partial_{\nu}	\omega^{a}	] \]
\[ +	{	\tilde	\x}^a	[Z^a	+	\e^{\m}	\pa_{\m}
\x^a	] \]
\be + {\tilde	Z}^a	[	c^{\a}
\s^{\mu}_{\a	\dot{\b}	}	\ov{c}^{\dot	\b}
\pa_{\m}	\x^a	+	\e^{\m}	\pa_{\m}	Z^a	] \Bigr
\}
\la{actsource}
\ee
Finally we come to the most important
part of the action for present purposes.  This is the
part $S_{\F}$  that introduces
the operator that may have a supersymmetry
anomaly.    It is not possible to write this
in a general way.  We refer the reader to
section (\ref{secexample}) for a specific example of
such an operator.

\section{Ward	Identity and Construction of BRS operator $\d$}

Once we have the action in the above form,
the Ward identity takes the Zinn-Justin \ci{zinn} `Master
Equation' form (a complete derivation of this for
D=4 super Yang-Mills can be found in \ci{d10}):
: \be \int	d^4	x	\Bigl	\{ \fr{
\d	\G}{	\d	X_i	} \fr{	\d	\G}{	\d	{\tilde
X}_i}
\Bigr \}
=0 \ee Here $X_i$ represents all the
fields and sources in the action. For one
loop amplitudes this reduces to the form \be
 \d	\G  =0 \ee where the BRS operator is
found from the action as follows: \be \d=
\int	d^4	x	\Bigl	\{ \fr{	\d	S_{\rm Total}}{	\d
X_i	} \fr{ \d	}{	\d	{\tilde	X}_i} + \fr{
\d	S_{\rm Total}}{	\d
{\tilde	X}_i} \fr{	\d	}{	\d	X_i	}
\Bigr \}
 \ee
Now the invariance of the action, and
hence the Master Equation, result  from the
invariance of the YM, chiral and Matter
actions , and from the nilpotence of the
transformations of the fields.   We can
recover these transformations as follows
from the action: \be \d	{ 	X_i}(x) =	\fr{\d
S_{\rm Total}}{\d	{\tilde X}_i(x)} \ee and the
transformations of the sources in $\d$ are
given by: \be
\d	{ 	{\tilde X}_i}(x) =
\fr{\d S_{\rm Total} }{\d	{  X}_i(x)}
\la{transsource}
\ee

\section{The	Full	BRS	Operator}
\la{full}
The following transformations are the gauge
and supersymmetry  transformations  for the
theory with spontaneous breaking of the
gauge symmetry.  They were used  in
$S_{\rm Ghost}$ and  $S_{\rm Sources}$ in the above
to derive the form of the action itself.
\begin{equation}	\delta	A^i	=	c^{\a}
\y^i_{\a} +	T^{ai}_j	\w^a	(m	\;	u^j	+	A^j) +
\epsilon^{\mu}	\partial_{\mu}	A^i
\la{deltaA} \end{equation}
 \begin{equation} \delta	\y^i_{\a}	=
\bl (D_{\m}	(m u + A) \br )^i \s^{\m}_{\a	\dot{\b}}
\overline{c}^{\dot{\b}} +	F^i	c_{\a} +
T^{ai}_j	\w^a	\y_{\a}^j +	\epsilon^{\mu}
\partial_{\mu}	\y_{\a}^i \la{deltapsi} \ee
\be \delta	F^i	= (
D_{\m}	\y)^{i	\a}	\s^{\m}_{\a	\dot{\b}}
\overline{c}^{\dot{\b}}	+	T^{ai}_j	\w^a	F^j
+	\epsilon^{\mu}	\pa_{\mu}	F^i
\la{deltaF}
\ee
\[	\delta V_{\mu}^{a}	=	D_{\mu}^{ab}
\omega^{b}	+ \fr{1}{2}	c^{\a}	\s^{\mu}_{\a
\dot{\b}	} \ov{\lambda}^{a	\dot{\b}	} \] \be
+	\fr{1}{2}	\l^{a	\a	} \s^{\mu}_{\a	\dot{\b}	}
\ov{c}^{\dot{\b}	} +	\epsilon^{\nu}
\partial_{\nu}	V_{\mu}^{a} \la{deltaV}
\end{equation}
\begin{equation} \delta	\lambda^{a}_{\a} =
\frac{1}{2}	G_{\mu\nu}^{a}	\s^{\mu	\nu}_{\a
\b}	c^{\b} -	f^{abc}	\lambda^{b}_{\a}
\omega^{c}	+	i	D^{a}	c_{\a} +	\epsilon^{\nu}
\partial_{\nu}	\lambda^{a}_{\a}
\end{equation}
\begin{equation} \delta	D^{a}	= \fr{-i}{2}
c^{\a}	\s^{\mu}_{\a	\dot{\b}	} D_{\m}^{ab}
\ov{\l}^{b	\dot{\b}	} +\fr{i}{2}	D_{\m}^{ab}
\l^{b	\a	} \s^{\mu}_{\a	\dot{\b}	}
\ov{c}^{\dot	\b} +	f^{abc}	D^{b}	\omega^{c} +
\epsilon^{\nu}	\partial_{\nu}	D^{a}
\end{equation}
 \begin{equation} \delta
\omega^{a}	=	-	\frac{1}{2}	f^{abc} \omega^{b}
\omega^{c} +	c^{\a}	\s^{\mu}_{\a	\dot{\b}	}
\ov{c}^{\dot	\b}	V_{\mu}^{a} +
\epsilon^{\nu}	\partial_{\nu}	\omega^{a}
\end{equation}
\begin{equation}
\delta	\x^a	=	Z^a	+	\e^{\m}	\pa_{\m}	\x^a
\la{deltaxi} \end{equation}
\begin{equation} \delta
Z^a=	c^{\a}	\s^{\mu}_{\a	\dot{\b}	}
\ov{c}^{\dot	\b} \pa_{\m}	\x^a	+	\e^{\m}
\pa_{\m}	Z^a \end{equation}
 \begin{equation} \delta
\epsilon_{\mu}	= -	c^{\a}	\s^{\mu}_{\a
\dot{\b}	}	\ov{c}^{\dot	\b} \end{equation}
\begin{equation} \delta	c^{\a}	=	0
\end{equation} \begin{equation} \delta
\ov{c}^{\dot	\a}	=	0 \end{equation}

The above transformations on the fields are
nilpotent--this is all that is needed to
derive the Master Equation, which then
implies that $\d$ is also nilpotent on the
sources.

Next we find the variations of the sources using (\ref{transsource})
and the foregoing expressions (\ref{stot}),
(\ref{actym}),
(\ref{actmat}), (\ref{actchiral}),
(\ref{actghost}) and
(\ref{actsource}) for the parts of the action:
\[ \delta	{\tilde
F}_i	= D^j_{\mu	i}	{\ov	D}_j^{\m	k	} (m	{\ov
u}_k	+	{\ov	A}_k) \] \[ +	{\ov	\y}^{\dot
\a}_{	j}	T^{aj}_{i}	{\ov	\l}^a_{\dot	\a	} +
D^a	T^{aj}_{i} (m	{\ov	u}_j	+	{\ov	A}_j) \] \[
+ 2 m	g_{ij}(u)	F^j
+ 3 g_{ijk}(u)	(	2	A^j	F^k +
\y^{\a	j}	\y_{\a}^k	)
+ \fr{12}{m} g_{ijkl}	(		A^j A^k	F^l +
\y^{\a	j}	\y_{\a}^k A^l	)
\] \[ + Z^{a} k	\;	m	\;
{\ov	T}^{aj}_i	{\ov	u}_j +	\xi^{a}	k	\;	m	\;
{\ov	T}^{aj}_k	{\ov	u}_j T^{ak}_i	\w^a \]
\be
+	T^{aj}_i	{\tilde	F}_j	\w^a +	(D_{\m}
{\tilde	\y})_{i}^{	\a}	\s^{\m}_{\a	\dot{\b}}
\overline{c}^{\dot{\b}}	+	\epsilon^{\mu}
\partial_{\mu} {\tilde	F}_i \la{deltatildeF}
\ee
\[
\delta	{\tilde	\y}_{i	\a}	=
	\sigma^{\m}_{\;\a	\dot	\b}	(D_{\mu}
\ov{\y}^{\dot	\b})_i +	{\ov T}^{aj}_{i}	\l^a_{\a}
(m	{\ov	u}_j	+	{\ov	A}_j) \] \[ + 2 m
{g}_{ij}(u)	\y^{\a	j}
+ 6 g_{ijk}(u)	A^j
\y_{\a}^k
+ \fr{12}{m} g_{ijkl}	A^j A^k
\y_{\a}^l
+	k	\;	m	\;	\x^a	{\ov	T}^{aj}_i
{\ov	u}_j c_{\a} \] \be +	{\tilde	F}_i	c_{\a}
+	T^{aj}_i	\w^a	{\tilde	\y}_{j	\a} +
	(D_{\m}	{\tilde	A})_i \s^{\m}_{\a	\dot{\b}}
\overline{c}^{\dot{\b}} +	\epsilon^{\mu}
\partial_{\mu}	{\tilde	\y}_{i	\a}
\la{deltatildepsi} \ee
\[ \delta	{\tilde	A}_i	= -	\ov{
F}_i +	2 m	g_{ij}(u)	A^j
+ 3  g_{ijk}(u)	A^j	A^k
+ \fr{4}{m} g_{ijkl}	A^j	A^k A^l \]
\be +c^{\a}	{\tilde	\y}_{i	\a} +	T^{aj}_i
\w^a	{\tilde	A}_j +	\epsilon^{\mu}
\partial_{\mu}	{\tilde	A}_i \la{deltatildeA}
\ee
\[
	\delta	{\tilde	V}^{\mu  a} =	(D^{\n}
G^{\mu	}_{\nu})^a  -	\frac{1}{2}	f^{abc}
\lambda^{b	\a} \s^{\m}_{\a	\dot{\b}	}
\ov{\lambda}^{b	\dot{\b}	} \] \[ +	T^{ai}_{j}
(m	u^j	+	A^j)	{\ov	D}_i^{\m	k	} (m	{\ov	u}_k
+	{\ov	A}_k) +	{	D}_i^{\m	k	}	(m	u^i	+	A^i)
{\ov	T}^{aj}_{k} (m	{\ov	u}_j	+	{\ov	A}_j) \]
\[ +	T^{ai}_{j} \y^{\a	j}	\sigma^{\m}_{\;\a
\dot	\b}	\ov{\y}^{\dot	\b}_i -
\partial^{\mu}	Z^{a} + T^{ai}_{j}	(m	\;	u^j	+
A^j	)	{\tilde	\y}_{i}^{	\a} \s^{\m}_{	\a
\dot{\b}}	\overline{c}^{\dot{\b}} + \]
\[ +
T^{ai}_{j}	{\tilde	A}_i	\y^{j	\a}
\s^{\m}_{\a	\dot{\b}}	\overline{c}^{\dot{\b}}
+ {\ov	T}^{aj}_{i}	{\tilde {\ov A}}_j
{\ov	\y}^{i	\dot	\a}	{\ov
\s}^{\m}_{	\dot	\a	\b}
	c ^{	\b	} \] \[ +	(D_{\n}	{\tilde
\lambda})^{	a	\a}
	\s^{\m \nu}_{ \a	\b}	c^{\b} +	(D_{\n}
{\tilde	{\ov	\l	}	})^{a	\dot	\a}
	\ov{\s}^{\m \nu}_{ \dot{\a}	\dot{\b}	}
\ov{c}^{\dot	\b} \]
\[ +	f^{abc} {\tilde
D}^{b}	[ \fr{-i}{2}	c^{\a}	\s^{\mu }_{\a
\dot{\b}	} \ov{\l}^{c	\dot{\b}	} +\fr{i}{2}
\l^{c	\a	} \s^{\mu}_{  \a	\dot{\b}	}
\ov{c}^{\dot	\b}	] \]
\be + f^{abc}(	{\tilde
V}^{\mu  b}	-	\partial^{\mu}
	\xi^{b}	)
	\omega^{c} +	{\tilde	\omega}^{a}
	c^{\a}	\s^{\mu}_{\a	\dot{\b}	}	\ov{c}^{\dot
\b} +	\epsilon^{\nu}	\partial_{\nu}	{\tilde
V}^{\mu a} \la{deltatildeV} \ee
\[ \delta	{\tilde
\lambda}^{a}_{\a} =
\fr{-1}{2}
\s^{\m}_{\a	\dot{\b}	}
(D_{\m}	{\ov	\l})^{a	\dot	\b} +	{\ov T}^{a	i}_j	(m
\;	{\ov	u}_i	+	A_i)	\y^{\a	j}
	-	\frac{1}{2}	({\tilde	V}^{a}_{\m}	-
\pa_{\m}	\x^a) \s^{\m}_{\a	\dot{\b}	}
\ov{c}^{	\dot{\b}	} \] \be +	f^{abc}	{\tilde
\lambda}^{b}_{\a}	\omega^{c} +
\fr{i}{2}
(D_{\m}
{\tilde	D})^{	a	} \s^{\mu	}_{\a	\dot	\b}	{\ov
c}^{\dot	\b} +	\epsilon^{\nu}	\partial_{\nu}
{\tilde	\lambda}^{a}_{\a} \ee
\[ \delta	{\tilde	D}^{a}	=
	D^{a}	+ T^{aj}_{i} (m	{\ov	u}_j	+	{\ov	A}_j)
(m	u^i	+	A^i) \] \be + i	{\tilde	\l}^{a	\a}
c_{\a} -i	{\tilde	{\ov	\l}}^{a	\dot	\a}
\ov{c}^{\dot	\a} +	f^{abc}	{\tilde	D}^{b}
\omega^{c} +	\epsilon^{\nu}	\partial_{\nu}
{\tilde	D}^{a} \ee
\[ \delta	{\tilde	\omega}^{a}	=
D^{\m	a	b}	({\tilde	V}_{\m}^b	-	\pa_{\m}
\x^a) +	\epsilon^{\nu}	\partial_{\nu}
{\tilde	\omega}^{a} \] \[ + \;	k	\;	m	\;
\xi^{b}
	T^{bj}_i	u^i T^{a	k}_j	(	m	\,	{\ov	u}_k	+
{\ov	A}_k) +\;	k	\;	m	\;	\xi^{b}	{\ov
T}^{bj}_i	{\ov	u}_j T^{ai}_j	(	m\;	u^j	+	A^j)
\] \[ +
[ {\tilde	F}_i T^{ai}_j	(	m	\;	u^j	+
A^j) +	{\tilde	\y}_i^{\a} T^{ai}_j	\y_{\a}^j
	+ {\tilde	A}_i
	T^{ai}_j	F^j
+{\rm comp.\; conj. } ]
\] \be
	-	f^{abc}	{\tilde	\lambda}^{b}_{\a}	\l^{\a
c} +	f^{abc}	{\tilde	{\ov	\l	}	}^{b	\dot	\a}
	\ov{\lambda}^{c}_{\dot	\a} + f^{abc}
{\tilde	D}^{b} D^{c} +	f^{abc}	{\tilde
\omega}^{b} \omega^{c} \ee
\[ \d	{\tilde
\x^a}	=	\fr{k}{2}
	c^{\a}	\s^{\mu}_{\a	\dot{\b}	}
\ov{c}^{\dot	\b} \pa_{\m}	\x^a
-	\pa_{\m}	[
D_{\mu}^{ab}	\omega^{b}	+ \fr{1}{2}	c^{\a}
\s^{\mu}_{\a	\dot{\b}	} \ov{\lambda}^{a
\dot{\b}	}
	+	\fr{1}{2}	\l^{a	\a	} \s^{\mu}_{\a
\dot{\b}	} \ov{c}^{\dot{\b}	}	] \] \[ +	k\;
m\;
	T^{ai}_{j}	u^j [	{\ov	c}^{\dot	\a}	{\ov
\y}_{	i	\dot	\a} +	{\ov	T}^{ak}_{i}	\w^a (m
{\ov	u}_k	+	{\ov	A}_k)	] \] \[ +	k\;	m\;	{\ov
T}^{ai}_{j}	{\ov	u}_i
	[	{	c}^{	\a}	{	\y}^j_{	\a} +	T^{aj}_{k}
\w^a
(m	{u}^k	+	{	A}^k)	]
	\] \be +	c^{\a}	\s^{\mu}_{\a	\dot{\b}	}
\ov{c}^{\dot	\b} \pa_{\m}	{\tilde	Z}^a +
\e^{\m}	\pa_{\m}	{\tilde	\x}^a
\la{deltatildexi} \ee
\[ \delta
{\tilde	Z}^a= k	Z^a	+	\pa^{\m}	V^a_{\m}	+ k\;
m\;	{T}^{ai}_{j}	{u}^j	{\ov	A}_i \] \be +
k\;m\; {\ov	T}^{ai}_{j}	{\ov	u}_i	A^j +	{\tilde
\x}^a +	\e^{\m}	\pa_{\m}	{\tilde	Z}^a
\la{deltatildeZ} \ee
\section{The `linear' part of 	$\d$}

We will find it useful to collect certain
terms in $\d$  when they have a common
heritage.  The entire operator can be
written in the form:
 \[ \d	=	c^{\a}	\L_{\a}	+	{\ov	c}^{\dot	\a}
{\ov	\L}_{\dot	\a} +	c^{\a}	{\ov	\na}_{\a}	+
{\ov	c}^{\dot	\a}	{	\na}_{\dot	\a} +	m	c^{\a}
{\ov	\S}_{\a}	+	m	{\ov	c}^{\dot	\a}	{
\S}_{\dot	\a} \]
\be +	\d_{	{\rm	\;
Kinetic	}	} +	m	\d_{	\rm		Mass}
+	\d_{	{\rm	Nonlinear		}	}
+ \e^{\m}	\pa_{\m}
	-	c^{\a}	\s^{\m}_{\a	\dot	\b}	{\ov	c}^{\dot
\b}	(\e^{\m})^{\dag} \ee
 where  we  take terms which are homogeneous in the fields
or homogeneous with one ghost $c$ or $\ov c$ for
our `linear' terms.  These decompose further into:
\be
	\L_{\a}	=
	\L_{\a	{\rm	\;	Matter}}	+	\L_{\a	{\rm	\;
Gauge}} +	{\tilde	\L}_{\a	{\rm	\;	Matter}}	+
{\tilde	\L}_{\a	{\rm	\;	Gauge}} \ee \be
\na_{\dot	\a} = \na_{\dot	\a	{\rm	\;	Matter}}
+	\na_{\dot	\a	{\rm	\;	Gauge}	} + {\tilde
\na}_{\dot	\a	{\rm	\;	Matter}}
+	{\tilde	  \na }_{ \dot 	\a	{\rm	\;	Gauge}	}
+	{\na }_{ 	\dot \a	{\rm	\;	Ghost}	}
	\ee \[
	\d_{	{\rm	\;		Kinetic	}	} =
	\d_{	{\rm	\;	Kinetic	\;	Matter	}	}
\] \be +	{\ov	\d}_{	{\rm	Kinetic
\;	Matter	}	}
+	\d_{	{\rm	 Kinetic \; Gauge	}	}
+	\d_{	{\rm	 Kinetic \; Ghost 	}	} \ee
\[
	\d_{	{\rm	\;		Mass	}	} =
	\d_{	{\rm	\;		Mass	\;	Matter	}	} +	{\ov	\d}_{	{\rm	\;		Mass	\;
Matter	}	}
+	\d_{	{\rm	\;	Mass	\; Gauge } }
\]
\[
+	\d_{	{\rm	\;	Mass\; Matter \ra Gauge } }
+	{\ov \d}_{	{\rm	\;	Mass\; Gauge \ra Matter  } }
\]
\be
+	\d_{	{\rm	\;	Mass\; Ghost    } }
+	\d_{	{\rm	\;	Mass\; Matter \ra  Ghost } }
+	\d_{	{\rm	\;	Mass\; Ghost \ra Matter  } }
\ee
It is convenient to introduce
the following abbreviations:
\be
T^{ai}	=
T^{ai}_j	u^j \ee
\be {\ov	T}^{a}_{i}	=
{\ov T}^{aj}_{i}	{\ov	u}_j
= { T}^{aj}_{i}	{\ov	u}_j
\ee
\be G^{ab}	= T^{ai}
{\ov	T}^{b}_{i}	+	T^{bi}	{\ov	T}^{a}_{i} \ee
Then the above operators can be written in the form:
\be
	\L_{\a	{\rm	\;Matter}}	= \int	d^4	x	\Bigl	\{
	\y^i_{\a}	\fr{\d	}{\d	A^i} +
	F^i	\fr{\d	}{\d	\y^{i	\a}	} \Bigr	\}
\la{lambdamatter} \ee \be
	{\tilde	\L}_{\a	{\rm	\;Matter}}	= \int	d^4
x	\Bigl	\{
	{\tilde	\y}_{i_	\a}	\fr{\d	}{\d	{\tilde
A}_i} +
	{\tilde	F}_i	\fr{\d	}{\d	{\tilde	\y}_{i
}^{a}	} \Bigr	\} \la{tildelambdamatter}\ee \be
	\na_{\dot	\a	{\rm	\;Matter}}	= \int	d^4	x
\Bigl	\{
	\pa_{\m}	\y^{i	\a}	\s^{\m}_{\a	\dot{\a}}
	\fr{\d	}{\d	F^i	} +	\pa_{\m}	A^{i	}
\s^{\m}_{\a	\dot{\a}}
	\fr{\d	}{\d	\y^{i}_{	\a}	} \Bigr	\}
\la{nablamatter} \ee \be
	{\tilde	\na}_{\dot	\a	{\rm	\;Matter}}	=
\int	d^4	x	\Bigl	\{
	\pa_{\m}	{\tilde	\y}_{i}^{	\a}	\s^{\m}_{\a
\dot{\a}}
	\fr{\d	}{\d	{\tilde	F}_i	} +	\pa_{\m}
{\tilde	A}_{i	}	\s^{\m}_{\a	\dot{\a}}
	\fr{\d	}{\d	{\tilde	\y}_{i	\a}	} \Bigr	\}
\la{tildenablamatter} \ee \be
	{	\L}_{	\a	{\rm	\;	Gauge}}	= \int	d^4	x
\Bigl	\{
	\fr{1}{2}	\s^{\m}_{\a	\dot	\b	} {\ov
\l}^a_{\dot	\b}
	\fr{\d	}{\d	V^{a	\m}	} +	i	D^a	\fr{\d	}{\d
\l^a_{	\a}	}
	\Bigr	\} \la{lambdagauge} \ee \be
	{\tilde	\L}_{	\a	{\rm	\;	Gauge}}	= \int	d^4
x	\Bigl	\{
	-	\fr{1}{2}	\s^{\m}_{\a	\dot{\b}}
	{\tilde	V}^{\m	a} \fr{\d	}{\d	{\tilde	{\ov
\l}}^a_{\dot	\b}	} +	i	{\tilde	\l}^a_{	\a}
\fr{\d	}{\d {\tilde	D}^a	}	\Bigr	\}
\la{tildelambdagauge} \ee \be
	{\ov	\na}_{	\a	{\rm	\;	Gauge}	}	= \int	d^4	x
\Bigl	\{
	\fr{1}{2}	\pa_{\m}	V^a_{\n}	\s^{\m	\n}_{\a
\b}
	{c}^{\b}	\fr{\d	}{\d	\l^{a}_{	\b}	} -
\fr{i}{2} \s^{\m}_{\a	\dot{\b}}	\pa_{\m}	{\ov	\l}^{a
\dot	\b}
	\fr{\d	}{\d	D^a	} \Bigr	\} \la{nablagauge}
\ee \be
	{\tilde	{\ov \na}}_{ \a	{\rm	\;	Gauge}	}	=
\int	d^4	x	\Bigl	\{
	\s^{\m	\n}_{\a	\b}	\pa_{\n}	{\tilde	\l}^{a
\b}
	\fr{\d	}{\d	{\tilde	V}^{\m a} } +
\s^{\m}_{\a	\dot{\b}}	\pa_{\m}
	{\tilde	D}^a
	\fr{\d	}{\d {\tilde{\ov	\l}}^{a}_{	\dot	\b}
} \Bigr	\} \ee
\be
	\d_{	{\rm	\;	Kinetic	\;	Matter	}	}
= \int	d^4	x	\Bigl	\{
	\Box	{\ov	A}_i
	\fr{\d	}{\d	{\tilde	F}_i	} +	\s^{\m}_{\a
\dot{\b}}	\pa_{\m}	{\ov	\y}^{\dot	\b}_i
	\fr{\d	}{\d	{\tilde	\y}^{	\a}_i	} -	{\ov	F}_i
	\fr{\d	}{\d	{\tilde	A}_i	}
	\Bigr	\} \ee \[
	\d_{	{\rm	Kinetic	\;	Gauge	}	} =
\int	d^4	x	\Bigl	\{
	(	\Box	V^{a	\m}	-	\pa^{\m}	\pa^{\n}
V^a_{\n})
	\fr{\d	}{\d	{\tilde	V}^{\m	a}	} \] \be +
\s^{\m}_{\a	\dot{\b}}	\pa_{\m}	{\ov	\l}^{a
\dot	\b}
	\fr{\d	}{\d	{\tilde	\l}^a_{	\a}	} +	{\ov
\s}^{\m}_{\dot	\a	{\b}}	\pa_{\m}	{	\l}^{	a
\b}
	\fr{\d	}{\d	{\tilde	{\ov	\l}}^a_{	\dot	\a}	}
+	D^a	\fr{\d	}{\d	{\tilde	{D}}^a	}
\Bigr \}
\ee
\be
	\d_{	\rm	Mass	\;	Matter} = \int
d^4	x
	2 g_{ij}	\Bigl	\{ F^j
	\fr{\d	}{\d	{\tilde	F}_i	}
	+	\y_{j	\a}
	\fr{\d	}{\d	{\tilde	\y}_{i	\a}	} +	A^j
	\fr{\d	}{\d	{\tilde	A}_{i}	} \Bigr	\} \ee \be
\d_{	\rm	\;		Mass	\;	Gauge}
	=	m \int	d^4	x
	G^{ab}	V^{b	\m}
	\fr{\d	}{\d	{\tilde	V}^{\m	a}	} \ee \[
	\d_{	\rm	\;	Mass \;	Matter	\ra
Gauge	} =	 \] \be \int	d^4	x \Bigl	\{
 	 	(\pa_{\m}	V^{\m	a}
	+	D^a)	{\ov T}^{a	}_i	]
	\fr{\d	}{\d	{\tilde	F}_i	}
	+	{\ov	T}^{a	}_i	\l^a_{\a}
	\fr{\d	}{\d	{\tilde	\y}_{i	\a}	}
	\Bigr	\} \ee  \[	\d_{	\rm	\;	Mass
\;	Gauge	\ra Matter	}	=	\int	d^4	x \Bigl	\{
	[T^{ai}	\pa^{\m}	{\ov	A}_i
	+	{\ov	T}^{a	}_i	\pa^{\m}	{A}^i	]
	\fr{\d	}{\d	{\tilde	V}^{\m	a}	} \] \be +
{\ov	T}^{b	}_i	\y^{j}_{\a}
	\fr{\d	}{\d	{\tilde	\l}^a_{	\a}	} +	T^{ai}
{\ov	\y}_{i	\dot	\a}
	\fr{\d	} {\d	{\tilde	{\ov	\l}}^a_{	\dot	\a}
} +	(	{\ov	T}^{a	}_i	A^i	+	T^{ai}
	{\ov	A}_i	)
	\fr{\d	}{\d	{\tilde	{D}}^a	} \Bigr	\} \ee
 Here are some ghost and gauge-fixing
dependent supersymmetry transformations: \be
	{\ov \S}_{\a}
=	\int	d^4	x \Bigl \{
k	{\ov	T}^{b	}_i	\y^{i	\a} \fr{\d	}{\d
{\tilde	\x}^{a}	}
+ k	\;	\x^a	{\ov	T}^{b	}_i
	\fr{\d	}{\d	{\tilde	\y}_{i	\a}	}
\Bigr \}
\ee
\be {\ov \na}_{\a\;	{\rm Ghost}} =	\int	d^4	x	\Bigl	\{
	\fr{1}{2}	\s^{\mu}_{\a	\dot{\b}	} \pa_{\m}
\ov{\lambda}^{a	\dot{\b}	}
	\fr{\d	}{\d	{\tilde	\x^a}	}
	+ \fr{1}{2} \s^{\m}_{\a	\dot{\b}}	\pa_{\m}	\x^{	a}
\fr{\d	}{\d	{\tilde	{\ov	\l}}^a_{\dot	\b}	}
\Bigr	\} \ee
Here is  a rather
amorphous collection of ghost and
gauge-fixing dependent linear terms:
\[
\d_{	{\rm		Kinetic	\;	Ghost}} = \int
d^4	x	\Bigl	\{
- \pa^{\m}	Z^a	\fr{\d	}{\d	{\tilde	V}^{\m	a}	}
-	\Box	\w^a	\fr{\d}{\d	{\tilde	\x}^a	} \] \[
+	(	k	Z^a	+	\pa^{\m}	V^a_{\m}
+	{\tilde	\x}^a
) \fr{\d}{\d	{\tilde	Z}^a	}
+ T^{ai}	\w^a	\fr{\d	}{\d	A^i	}
+ {\ov T}^a_i	\w^a	\fr{\d	}{\d	{\ov A}_i	}
 \]
\be +	\pa_{\m}
\w^a \fr{\d	}{\d	V^a_{\m}	} +	\pa_{\m}	(
{\tilde	V}^{a	\m}	-	\pa^{\m}	\x^a	) \fr{\d
}{\d	{\tilde	\w}^a	}
+	Z^a
	\fr{\d	}{\d	\x^a	}
	\Bigr	\} \ee
\be \d_{	\rm	\;	Mass	\; Ghost}
	=	\int	d^4	x
	G^{ab} \Bigl	\{ k	\;	\w^b
	\fr{\d	}{\d	{\tilde	\x}^a	} +	k	\x^b
	\fr{\d	}{\d	{\tilde	\w}^a	} \Bigr	\} \ee \be
\d_{\rm Mass\;	Matter \ra Ghost} =	k	Z^a	 	{\ov
T}^{a	}_i
	\fr{\d	}{\d	{\tilde	F}_i	} \ee \[	\d_{	\rm
	Mass;	\;	Ghost	\ra Matter	} = \]
\be
	\int	d^4	x \Bigl	\{
	k	(	T^{ai}	{\ov	A}_i +	{\ov	T}^{a	}_i	A^i	)
	\fr{\d	}{\d	{\tilde	Z}^a	} +	(	T^{ai}
{\tilde	F}_i +	{\ov	T}^{b	}_i	{\tilde	{\ov
F}}	)
	\fr{\d	}{\d	{\tilde	\w}^a	} \Bigr	\} \ee

Finally  there are a large number  of terms
that are non-linear which  can all be deduced
given the total form of $\d$ above.
Unfortunately it is necessary to write them
all down and include them in the grading of
$\d$ even though most of them will play no
role in the spectral sequence that we will use
to find the cohomology.  The problem
is that only by going through all the steps
can one be sure that nothing important has
been missed.  In fact some of the non-linear
terms do play an important role, even though
most of them do not.

\section{Non-linear terms}

The non-linear terms in the above BRS transformations are:
\begin{equation}	\delta_{\rm Nonlinear}	A^i
=
 	T^{ai}_j	\w^a	 	A^j
 \la{deltanonlinA} \end{equation}
\[
\delta_{\rm Nonlinear}	{\tilde	F}_i	=
V^j_{\mu	i}	{ \pa}^{\m		}
 	{\ov	A}_j  +\pa_{\mu	 }[	{\ov	V}_i^{\m	k	}
 	  {\ov	A}_k ] +V^j_{\mu	i} {\ov	V}_j^{\m	k	}
 	  (m	{\ov	u}_k	+	{\ov	A}_k) \] \[ +	{\ov
\y}^{\dot	\a}_{	j}	T^{aj}_{i}	{\ov
\l}^a_{\dot	\a	} + D^a	T^{aj}_{i}
 	{\ov	A}_j  \] \[
 + 3 g_{ijk}(u)
 	[	2	A^j	F^k +	\y^{\a	j}	\y_{\a}^k]
+ \fr{12}{m}  g_{ijkl}
 	[	A^j	A^k F^l +A^j 	\y^{\a	k}	\y_{\a}^l ]
 \] \[ +	\xi^{a}	k	\;
	m	\;	{\ov	T}^{aj}_k
{\ov	u}_j T^{ak}_i	\w^a \] \be +	T^{ak}_i
{\tilde	F}_j	\w^a +	(V_{\m}	{\tilde
\y})_{i}^{	\a}	\s^{\m}_{\a	\dot{\b}}
\overline{c}^{\dot{\b}}
\la{deltanonlintildeF} \ee \begin{equation}
\delta_{\rm Nonlinear}	\y^i_{\a}	=	(V_{\m}
A)^i \s^{\m}_{\a	\dot{\b}}
\overline{c}^{\dot{\b}} +	T^{ai}_j	\w^a
\y_{\a}^j \la{deltanonlinpsi} \ee \[
\delta_{\rm Nonlinear}	{\tilde	\y}_{i	\a}	=
	\sigma^{\m}_{\;\a	\dot	\b}	(V_{\mu}
\ov{\y}^{\dot	\b})_i +	T^{ai}_{j}	\l^a_{\a}
 	{\ov	A}_i  \] \[
+6 g_{ijk}(u) 	A^j	\y_{\a}^k
+ \fr{12}{m} g_{ijkl}	A^j A^k \y_{\a}^l
\]
\be +	T^{aj}_i	\w^a	{\tilde	\y}_{j	\a} +
	(V_{\m}	{\tilde	A})_i \s^{\m}_{\a	\dot{\b}}
\overline{c}^{\dot{\b}}
\la{deltanonlintildepsi} \ee
\be \delta_{\rm
Nonlinear} F^i	= (	V_{\m}	\y)^{i	\a}
\s^{\m}_{\a	\dot{\b}}
\overline{c}^{\dot{\b}}	+	T^{ai}_j	\w^a	F^j
\la{deltanonlinF} \ee
\[ \delta_{\rm
Nonlinear}	{\tilde	A}_i	= 3 g_{ijk}(u)	A^j	A^k
+ \fr{4}{m} g_{ijkl}	A^j A^k A^l
\]
\be +	T^{aj}_i	\w^a	{\tilde	A}_j
\la{deltanonlintildeA}
\ee \be
	\delta_{\rm Nonlinear} V_{\mu}^{a}	=
V_{\mu}^{ab}	\omega^{b}
\la{deltanonlinv}
\end{equation}
\[
		\delta_{\rm Nonlinear}
	{\tilde	V}^{\mu a} =	  f^{abc} V^{b \n}	[
\pa^{\mu}	V_{\nu}^{c} - \pa_{\nu}
V^{\mu c} ]  + f^{abc} \pa^{\nu} [
V^{\m b}	V_{\nu}^{c} ] +
 f^{abc}  V^{b \n} f^{cde}  V^{\m d}
V_{\nu}^{e}
\]
\[  -	\frac{1}{2}	f^{abc}
\lambda^{b	\a} \s^{\m}_{\a	\dot{\b}	}
\ov{\lambda}^{b	\dot{\b}	} \]
\[ +	T^{ai}_{j}	 A^j 	\pa^{\m}    {\ov	A}_i +
{\ov T}^{ai}_{j}	 {\ov	A}_i \pa^{\m}
A^j 	  	   +	T^{ai}_{j}	 m	u^j	 	{\ov
V}_i^{\m	k	}
 {\ov	A}_k  +	{	V}_i^{\m	k	}	m	u^i		{\ov
T}^{aj}_{k}
 	{\ov	A}_j  \] \[ +	T^{ai}_{j}	 	A^j  {\ov
V}_i^{\m	k	}
 m	{\ov	u}_k	  +	{	V}_i^{\m	k	}	 	A^i {\ov
T}^{aj}_{k}
   m	{\ov	u}_j	   +	T^{ai}_{j}	 	A^j 	{\ov
V}_i^{\m	k	}
 	{\ov	A}_k  +	{	V}_i^{\m	k	}	 	A^i 	{\ov
T}^{aj}_{k}
 	{\ov	A}_j
\]
\[
+	T^{ai}_{j} \y^{\a	j}
\sigma^{\m}_{\;\a	\dot	\b}	\ov{\y}^{\dot
\b}_i
+ T^{ai}_{j}	 	A^j	 	{\tilde	\y}_{i}^{
\a} \s^{\m}_{	\a	\dot{\b}}
\overline{c}^{\dot{\b}}
+ {\ov T}^{ai}_{j}	 	{\ov A}_i  	{\ov {\tilde	\y}}^{j \dot \a}
{\ov \s}^{\m}_{	\dot \a
 \b }  {c}^{ \b }
\]
\[
+ T^{ai}_{j}
{\tilde A}_i  	 	{ \y}^{j  \a} \s^{\m}_{	\a	\dot{\b}}
\overline{c}^{\dot{\b}}
+ {\ov T}^{ai}_{j}	 	{\tilde {\ov A}}^j  	{\ov { 	\y}}_i^{\dot \a}
{\ov \s}^{\m}_{	\dot \a \b }  {c}^{ \b }
\]
\[
+	(V_{\n}	{\tilde
\lambda})^{	a	\a}
	\s^{\mu	\nu}_{\a	\b}	c^{\b} +	(V_{\n}
{\tilde	{\ov	\l	}	})^{a	\dot	\a}
	\ov{\s}^{\mu	\nu}_{\dot{\a}	\dot{\b}	}
\ov{c}^{\dot	\b}
\]
 \[
+	f^{abc} {\tilde
D}^{b}	[ \fr{-i}{2}	c^{\a}	\s^{\mu}_{\a
\dot{\b}	} \ov{\l}^{c	\dot{\b}	} +\fr{i}{2}
\l^{c	\a	} \s^{\mu}_{\a	\dot{\b}	}
\ov{c}^{\dot	\b}	]
\]
\be
+ f^{abc}(	{\tilde
V}^{\mu b}	-	\partial^{\mu}
	\xi^{b}	)
	\omega^{c} +	{\tilde	\omega}^{a}
	c^{\a}	\s^{\mu}_{\a	\dot{\b}	}	\ov{c}^{\dot
\b} \la{deltanonlintildeV}
\ee
\begin{equation} \delta_{\rm Nonlinear}
\lambda^{a}_{\a} =	\frac{1}{2}	f^{abc}
V^b_{\mu} V^c_{\nu} 	\s^{\mu	\nu}_{\a	\b}
c^{\b} -	f^{abc}	\lambda^{b}_{\a}
\omega^{c}	  \end{equation} \[
\delta_{\rm
Nonlinear}		{\tilde	\lambda}^{a}_{\a} = \fr{1}{2}
\s^{\m}_{\a	\dot{\b}	}	(V_{\m}	{\ov	\l})^{a
\dot	\b} +	T^{a	i}_j	 	A_i 	\y^{\a	j} \] \be
+	f^{abc}	{\tilde	\lambda}^{b}_{\a}
\omega^{c} +
\fr{i}{2}
(V_{\m}	{\tilde	D})^{	a	} \s^{\mu
}_{\a	\dot	\b}	{\ov	c}^{\dot	\b} \ee
\begin{equation} \delta_{\rm Nonlinear}
D^{a}	= \fr{-i}{2}	c^{\a}	\s^{\mu}_{\a
\dot{\b}	} V_{\m}^{ab}	\ov{\l}^{b	\dot{\b}	}
+\fr{i}{2}	V_{\m}^{ab}	\l^{b	\a	}
\s^{\mu}_{\a	\dot{\b}	} \ov{c}^{\dot	\b} +
f^{abc}	D^{b}	\omega^{c} \end{equation} \[
\delta_{\rm Nonlinear}	{\tilde	D}^{a}	=
T^{aj}_{i}
 {\ov	A}_j
 	A^i \] \be +	f^{abc}	{\tilde	D}^{b}
\omega^{c} \ee \begin{equation} \delta_{\rm
Nonlinear}	\omega^{a}	=	-	\frac{1}{2}	f^{abc}
\omega^{b}	\omega^{c} +	c^{\a}	\s^{\mu}_{\a
\dot{\b}	}	\ov{c}^{\dot	\b}	V_{\mu}^{a}
\end{equation}
\[
\delta_{\rm Nonlinear}
{\tilde	\omega}^{a}	= V^{\m	a	b}	({\tilde
V}_{\m}^b	-	\pa_{\m}	\x^a) \] \[ + \;	k	\;	m
\;	\xi^{b}
	T^{bj}_i	u^i T^{a	k}_j	 	{\ov	A}_k  +\;	k
\;	m	\;	\xi^{b}	{\ov	T}^{bj}_i	{\ov	u}_j
T^{ai}_j	 	A^j \] \[
+	[ {\tilde	F}_i
T^{ai}_j	 	A^j  +	{\tilde	\y}_i^{\a}
T^{ai}_j	\y_{\a}^j
	+ {\tilde	A}_i
	T^{ai}_j	F^j  \; + {\rm comp. \; conj.} ]
\]
\be
	-	f^{abc}	{\tilde	\lambda}^{b}_{\a}	\l^{\a
c} +	f^{abc}	{\tilde	{\ov	\l	}	}^{b	\dot	\a}
	\ov{\lambda}^{c}_{\dot	\a} + f^{abc}
{\tilde	D}^{b} D^{c} +	f^{abc}	{\tilde
\omega}^{b} \omega^{c} \la{deltanonlintildom}
\ee \begin{equation} \delta_{\rm Nonlinear}
\x^a	=	0  \la{deltanonlinxi} \end{equation} \[
\delta_{\rm Nonlinear}	{\tilde	\x^a}	=	 -
\pa_{\m}	[	V_{\mu}^{ab}	\omega^{b}]
	 \] \[ +
k\;	m\;
	T^{ai}_{j}	u^j {\ov	T}^{ak}_{i}	\w^a
 	{\ov	A}_k	  \] \[ +	k\;	m\;	{\ov
T}^{ai}_{j}	{\ov	u}_i
	T^{aj}_{k}	\w^a {	A}^k
	\] \be +	c^{\a}	\s^{\mu}_{\a	\dot{\b}	}
\ov{c}^{\dot	\b} \pa_{\m}	{\tilde	Z}^a
\la{deltanonlintildexi} \ee \begin{equation}
\delta_{\rm Nonlinear}	Z^a =	c^{\a}
\s^{\mu}_{\a	\dot{\b}	}	\ov{c}^{\dot	\b}
\pa_{\m}	{	\x}^a \end{equation} \[
\delta_{\rm Nonlinear}	{\tilde	Z}^a= 0
\la{deltanonlintildeZ} \] \begin{equation}
\delta_{\rm Nonlinear} \epsilon_{\mu}	= -
c^{\a}	\s^{\mu}_{\a	\dot{\b}	}	\ov{c}^{\dot
\b} \end{equation}

\section{Gradings}

We want to examine some aspects of the BRS
cohomology of the foregoing formidable
operator $\d$.  To do this we shall use a
spectral sequence, which in turn is
generated by a grading.  Familiarity with the methods and
results of \ci{cmp1} \ci{cmp2} \ci{cqg} \ci{prl}  \ci{dm} and
\ci{dmr} will be assumed here.  See also
\ci{bdk} \ci{b1} \ci{b2} for a different approach.
  The choice of the grading is far from unique.  Each grading
has some advantages and some disadvantages.
To find the present one, all
the terms in $\d$ were put on a microcomputer
spreadsheet and a number of
possibilities were tried.  For this purpose, $\d$ was divided into
97 different terms. In general, for an
arbitrary (integral) grading, $\d$ will
split up into a sum of the form
 \be \d	=	\sum_{i=- N_{-} }^{  N_{+} }\d_i
\ee
 However a spectral sequence arises in the
manner contemplated in  \ci{cmp1} only when
the lower limit satisfies $N_- = 0$.
We will call this a positive grading.
  By experimenting  on the
spreadsheet one quickly finds a number of
gradings that grade $\d$ positively, and many more that give
$N_- < 0$. It is necessary to try to choose one of
the positive gradings that yields a sufficiently simple
form for the low $\d_i$ and their
Laplacians  $\D_i$.  There is often an
`equivalence class of gradings' which all
give rise to the same $E_{\infty}$, and, in
such cases, it does not make much
difference  which grading in the class one
chooses.

A grading that seems very  useful  for the
present problem is: \[ N_{\rm Grading} =
3	[
N'_{\rm	Matter}	 +	{\tilde	N'}_{\rm
{Matter}}   +	{\ov	N'}_{\rm	Matter} +
{\tilde	{\ov	N'}}_{\rm	{Matter}} ] \]
\[ +	2[
N'_{\rm	Gauge}+	{\tilde	N'}_{\rm	{Gauge}} ]
+7 [N_{\rm	Matter} + {\ov N}_{\rm	Matter}	 ]
+ 11 N_{\rm	Gauge} \] \[ + 	3[  N(c) +
N({\ov c}) ] +	N_{\rm	Gauge	Fixing}
+	4  {\tilde N}_{\rm	Gauge	Fixing} \]
\be
+ N(\e)+ 17 N(\w) + 2 N({\tilde \w})  \ee where \be N_{\rm	Matter}	=
	N[A]	+	N[\y]	+	N[F] \ee \be {\tilde	N}_{\rm
{Matter}} = N[{\tilde	A}] +	N[{\tilde	\y}] +
N[{\tilde	F}] \ee \be N'_{\rm	Matter}	= 3
N[A]	+	2	N[\y]	+	N[F] \ee \be {\tilde
N'}_{\rm	{Matter}} = 3	N[{\tilde	A}] +	2
N[{\tilde	\y}] +	N[{\tilde	F}] \ee \be N_{\rm
Gauge} = N[V]	+	N[\l]	+	N[{\ov	\l}]	+	N[D] \ee
\be {\tilde	N}_{\rm	{Gauge}} = N[{\tilde	V}]
+	N[{\tilde	\l}]	+	N[{\tilde	{\ov	\l}}]	+
N[{\tilde	D}] \ee \be N'_{\rm	Gauge} = 3	N[V]
+	2	N[\l]	+	2	N[{\ov	\l}]	+	N[D] \ee \be
{\tilde	N'}_{\rm	{Gauge}} = N[{\tilde	V}]	+	2
N[{\tilde	\l}]	+	2	N[{\tilde	{\ov	\l}}]	+	3
N[{\tilde	D}] \ee
\be N_{\rm	Gauge	Fixing} =
N[Z]   +	N[\x]
\ee
\be {\tilde N}_{\rm	Gauge	Fixing} =
 	N[{\tilde	Z}]   +	N[{\tilde	\x}]
\ee
These definitions are motivated by the simple relations:
\be
[ N'_{ {\rm Matter} } ,\L_{\a {\rm Matter} } ]
=
- \L_{\a {\rm Matter}} ]
\ee
\be
[ N'_{{\rm Matter}} ,\na_{\dot \a {\rm Matter}} ]
=
\na_{\dot \a {\rm Matter} } ]
\ee
with similar relations for all the other cases.
This grading was found by trying to duplicate
the success of the grading used for the pure chiral
case without gauge fields or sources for BRS
variations. It is also adapted for separating the
gauge fields from the matter fields, since mixing them
causes difficulties.  It would be nice to treat the
chiral fields and sources in exactly the same way,
but this is incompatible with a positive grading and
the other requirements.  It seems a good idea to
keep the degrees of $V$ and $\w$ identical in view of
the BRS cohomology of pure Yang-Mills, so that
the operator $\pa_{\m} \w^a \fr{\d}{\d V^a_{\m} }$
occurs in $\d_0$ and so eliminates all derived
$\w$ fields.  Next  one wants to ensure that the
operator $\e^{\m}  \pa_{\m}$ occurs after the operator
$\pa_{\m} \w^a \fr{\d}{\d V^a_{\m} } $ so that the
results of \ci{cmp1} can eventually be used.  The
operator $c^{\a}
\s^{\m}_{\a \dot \b}
{\ov c}^{\dot \b} (\e^{\m})^{\dag}$ must be of higher
order than $\e^{\m}  \pa_{\m}$ or else there  are
very difficult mixings.

Once these criteria are met, there is very little
freedom in choosing the grading left in the problem.
The other terms are determined so that $\d_0$
generates some strong and simple restrictions.

Actually, it is clear that a great deal of
information is obtained from the existence of a
grading like the present one. It  is
susceptible of much more exploitation.
I believe that the full problem can also be
solved using this or a similar grading, but it
requires lots more work, and there are still some tricky
problems.

We note that the ghost number is given by
the grading operator \[ N_{\rm	Ghosts} =
N[\w]	+	N[\e]	+	N(c)	+	N[	{\ov	c}] -	N[\x] \]
\[ -	N[{\tilde	V}] -	N[{\tilde	\y}]
	-	N[{\tilde	{\ov	\y}}] -	2	N[{\tilde	\w}] -
N[{\tilde	\l}]
	-	N[{\tilde	{\ov	\l}}] \] \be
	-	N[{\tilde	{	F}}] -	N[{\tilde	{\ov	F}}]
	-	N[{\tilde	{	A}}] -	N[{\tilde	{\ov	A}}]
	-	N[{\ti {	D}}]
	-	N[{\ti {	Z}}] \ee
 It satisfies the simple relations \be [
N_{\rm	Ghosts}, \d] = \d \ee \be [ N_{\rm
Ghosts}, S_{\rm Total}] = 0 \ee

\section{Space	$E_1$} Using the grading above,
we find
\be
\d = \sum_{i=0}^{i=50}
\d_i
\ee
and all these $\d_i$ (some of which are zero) will
now be presented and discussed in the context of the
limited result that we want to establish in this paper.
The first operator in the series is:
\[ \d_0	=
c^{\a}	\L_{\a \;{\rm Matter}} + {\ov	c}^{\dot
\a}	{\ov	\L}_{\dot	\a  \;{\rm Matter}} +
c^{\a}	{\tilde \L}_{\a\;{\rm Matter}} + {\ov
c}^{\dot	\a}
{\tilde
{ \ov	\L}}_{\dot \a  \;  {\rm Matter}} \]
\be + \int d^4 x \{ Z^a
\fr{\d}{\d \x^a}  +
{\tilde \x}^a \fr{\d}{\d {\tilde Z}^a}
+   \pa_{\m} \w^a
\fr{\d}{\d V_{\m}^a}
+  \pa_{\m} {\ti V}^{\m a}
\fr{\d}{\d {\ti\w}^a} \}
\ee

The cohomology of the c-dependent parts
of this operator are  already
known and have been analyzed at length in
\ci{cmp2}.  The next two terms simply eliminate
all  dependence on the four fields $\x, Z,
{\tilde \x}$ and ${\tilde Z}$ from the
cohomology space $E_1$.  Their presence here means
that we can now also ignore all the terms in
the higher operators $\d$ that depend on
these fields, since these will all give zero
when sandwiched between projection operators
onto the space $E_1$. The next two terms are also simple to
analyze--see \ci{cmp1}.
  In summary, the equations determining the space $E_1$
are:
\be (\w^a_{\m_1 \cdots \m_k})^{\dag} E_1 = 0
;\; ( k \geq 1)
\ee
\be (V^a_{(\m_1, \cdots \m_k)})^{\dag} E_1 = 0
;\; ( k \geq 1)
\la{vequation}
\ee
\be ({\ti \w}^a_{\m_1 \cdots \m_k})^{\dag} E_1 = 0
;\; ( k \geq 0)\ee
\be ({\ti V}^{a
 \m}_{\;\;\;\; , \m \m_1  \cdots \m_k})^{\dag} E_1 = 0
;\; ( k \geq 0)
\ee
\be ({\ti \x}^a_{\m_1 \cdots \m_k})^{\dag} E_1 = 0
;\; ( k \geq 0)
\ee
\be ({\ti Z}^a_{\m_1 \cdots \m_k})^{\dag} E_1 = 0
;\; ( k \geq 0)
\ee
\be ({ \x}^a_{\m_1 \cdots \m_k})^{\dag} E_1 = 0
;\; ( k \geq 0)
\ee
\be ({ Z}^a_{\m_1 \cdots \m_k})^{\dag} E_1 = 0
;\; ( k \geq 0)
\ee
\be
 \Bl [ \L_{\a \; {\rm Matter}} + {\ti \L}_{\a\; {\rm Matter}
}	\Br ] E_1	=	0
\la{lam}
\ee
\be
\Bl [
\ov{\L}_{\dot	\a\; {\rm Matter} }	 +
{\ti {\ov \L}}_{\dot	\a \; {\rm Matter} }	 \Br  ] E_1	=0
\la{lambar}
\ee
 \be
N(c) [	N_{\rm Matter}	+ {\ti N}_{ {\rm Matter}  } ]
E_1	=	0
\ee
\be
N({\ov{c}})	\,	[ \ov{N}_{ \rm Matter}
 + {\ti {\ov N}}_{ \rm Matter}]	E_1	=	0
\la{lam5}
\ee

We must now refer to \ci{dmr} for a discussion of the solution of the
equations involving $\L$. The current problem is no different
 except that there are additional  variables $\ti A$ etc.

\section{The Operator $d_1$}

The next operator in the sequence is:
\[ \d_1	= c^{\a}	\L_{\a\;{\rm
Gauge}} + {\ov	c}^{\dot	\a}	{\ov	\L}_{\dot
\a\;{\rm Gauge}} + c^{\a}	{\tilde
\L}_{\a\;{\rm Gauge}} + {\ov	c}^{\dot	\a}
{\tilde { \ov	\L} }_{\dot	\a\;{\rm Gauge}}
+ \e^{\m} \pa_{\m}
\]
\[+ \int d^4 x
\Bl \{
 {\ov F}_i
\fr{\d}{\d {\tilde A}_i}
+{  F}^i
\fr{\d}{\d {\tilde {\ov A}}^i}
+ m T^{ai} Z^a
\fr{\d}{\d {\ti {\ov F}}^i}
  + m {\ov T}^{a}_{i} Z^a
\fr{\d}{\d {\ti  F}_i}
\]
\[
 + m {\ov T}^{a}_{i} c^{\a} \x^a
\fr{\d}{\d {\ti \y }^{\a}_i}
 + m {  T}^{a i} {\ov c}^{\dot \a} \x^a
\fr{\d}{\d {\ti {\ov \y} }^{\dot \a i} }
\]
\[ + m {\ov   T}^{a}_i {\ti {\ov F}}^i
\fr{\d}{\d {\tilde \w}^a}
+ m  {  T}^{a i}  {\ti   F}_i
\fr{\d}{\d {\tilde \w}^a }
\]
\be + m T^{ai} \w^a
\fr{\d}{\d A^i}
 + m {\ov T}^{a}_{i} \w^a
\fr{\d}{\d {\ov A}_i }
\Br \}
\ee
Using the properties of $\P_1$ above, we immediately deduce that
$d_1$ has the considerably simpler form
\[
d_1 = \P_1
 \d_1	\P_1
= \P_1
\Bl \{
c^{\a}	\L_{\a\;{\rm
Gauge}} + {\ov	c}^{\dot	\a}	{\ov	\L}_{\dot
\a\;{\rm Gauge}} + c^{\a}	{\tilde
\L}_{\a\;{\rm Gauge}} + {\ov	c}^{\dot	\a}
{\tilde { \ov	\L} }_{\dot	\a\;{\rm Gauge}}
+ \e^{\m} \pa_{\m}
\]
\be
+ \int d^4 x \{ {\ov F}_i
\fr{\d}{\d {\tilde A}_i}
+{  F}^i
\fr{\d}{\d {\tilde {\ov A}}^i}
+ m T^{ai} \w^a
\fr{\d}{\d A^i}
 + m {\ov T}^{a}_{i} \w^a
\fr{\d}{\d {\ov A}_i }
\Br \} \P_1
\la{d1}
\ee

 Here we shall
not take up the large task of analyzing the
cohomology of these   $\L_{\a\;{\rm
Gauge}}$ operators. It looks  likely that
  this may  be done in a
fairly straightforward way along the lines
of \ci{dmr} and it also  appears that
the answer may again be highly nontrivial.

\section{Masses and the Equation of Motion}
\la{masssec}

In this paper we often have to deal with the dimensional
parameter $m$.  All other parameters can be chosen to be
dimensionless multiples of this parameter to various
powers.

Let us examine a very simple example.
Suppose that we start
with the simplest example of (\ref{action})
\be
S_{\F} = \int d^6 {\ov z}
\;
\fr{1}{m} \F^{\a} D^2 (S_1 D_{\a} S_2 )
=
\int d^4 x \;
\fr{1}{m} \{ \c^{\a} F_1 \y_{2 \a} + \cdots \}
\la{opstart}
\ee
and somehow generate the corresponding general form of (\ref{anom})
(We will ignore the index on $S_i$):
\[
\d \G_{\F} = \int d^6 {\ov z}
\;
\fr{1}{m} \F^{\a} c_{\a}
[g_1  m^3 {\ov S} + g_2 m^2 {\ov
S}^2 + g_3  m {\ov S}^3 + g_4   {\ov S}^4 ]
\]
\be
=
\int d^4 x \;
\fr{1}{m} \{ \c^{\a} c_{\a} [g_1  m^3 {\ov A} + g_2 m^2 {\ov
A}^2+ + g_3  m {\ov A}^3 + g_4   {\ov A}^4 ]
+ \cdots \}
\la{opstartanom}
\ee
Here the $\fr{1}{m} $ is inserted so that $\c^{\a}$
will have its canonical dimension of  $\fr{3}{2} $,
and the dimensions of the anomalous part are
fixed by simple dimensional counting.  All the
coefficients $g_i$ are dimensionless.
Then clearly some of the terms (\ref{opstartanom})
are related by the pure chiral part of the equation of
motion.  How do we pick out the physically important
part that remains in the cohomology space
when the equation of motion part is included?

The first question is whether $m$ should
be treated as a spacetime independent field
like the supersymmetry ghost $c_{\a}$ or merely
as a constant, for the  purposes of finding the BRS
cohomology of the operator $\d$. In fact, it is clearly necessary to
treat it as a field for dimensional consistency.   We will
now see how this works.

Suppose that  the $A$ field bit  of the pure chiral part of the
equation of motion in $\d$ for this case has the very simple form:
\be
\d = [ m A + g A^2 ] {\tilde A}^{\dag}
\ee
Then its adjoint is:
\be
\d =  {\tilde A} [ m A + g  A^2 ]^{\dag} =
{\tilde A} [  A^{\dag} m^{\dag} +g  A^{\dag} A^{\dag} ]
\ee
and we use the  relation
\be
[m^{\dag}, m]=1
\ee
thus treating $m$ like a constant field, rather than as a constant.

How does the cohomology work for this case?
Let us use the spectral sequence method for this
operator, using the grading
\be
N_{\rm Grading} = A A^{\dag}
+ {\tilde A} {\tilde A}^{\dag}
\ee
Then
  \be
\d_0 =   m \W
\ee
where
\be
\W =
A {\tilde A}^{\dag}
\ee
and
\be
\d_1 = g A^2 {\tilde A}^{\dag}
\ee
We easily see that
\be
\D_0 = [N(A) + N({\tilde A} ) ] N(m)
+ \W^{\dag} \W
\ee
so that
\be
E_1 = X(A) + Y(m)
\ee
where $X$ and $Y$ are arbitrary functions of the indicated
variables and no others.
Now clearly
\be
\d_1 E_1 = 0
\ee
because
$E_1$ is independent of $\tilde A$ and we have
\be
E_{\infty} = E_1
\ee
for the same reason: all the operators $d_r$ are zero
for $r \geq 1$ because they all need $\d_1 \P_r = 0$
in their construction.

But how do we find the correspondence
\be
E_{\infty} \ra H
\la{iso}
\ee
for use in extracting the physically
meaningful part of (\ref{opstartanom}) that is
in the cohomology space of the full operator $\d$
including the equation of motion term?

One way to see it is to write the above expression as
an expression which vanishes by the equation of motion
plus a remainder.  This is
\[
\int d^4 x \;
\fr{1}{m} \{ \c^{\a} c_{\a} [g_1  m^3 {\ov A} + g_2 m^2 {\ov
A}^2+ + g_3  m {\ov A}^3 + g_4   {\ov A}^4 ]
+ \cdots \}
\]
\[
= \int d^4 x \;
\fr{1}{m} \Bigl \{ \c^{\a} c_{\a}  \Bigl   [
[  m {\ov A}  + g  {\ov A}^2 ]
  [g'_1  m^2 +  g'_2  m {\ov A}
+  g'_3   {\ov A}^2 \Bigr ]
+ g'_4   {\ov A}^4
+ \cdots \}
\]
\be
=
\int d^4 x \;
\fr{1}{m} \Bigl \{ \c^{\a} c_{\a}  \Bigl   [
\d {\tilde A}
  [g'_1  m^2 +  g'_2  m {\ov A}
+  g'_3   {\ov A}^2 \Bigr ]
+ g'_4   {\ov A}^4
+ \cdots \}
\ee
and the coefficient of the last term, which is the anomaly,
is easily seen to be
\be
g'_4 = g_4 - g g_3 + g^2 g_2 - g^3 g_1
\la{g4}
\ee
This should be gauge-invariant and physically
meaningful.  Clearly the correspondence (\ref{iso})
here is simply an identity--we take the term
that has no masses to be our cohomology space and
all the others are in the image of $\d$, but some
care is needed to get the coefficient right as shown
above.   Let us check that this
coefficient is indeed singled out by the cohomology by
writing the expression in a different way:
\[
\int d^4 x \;
\fr{1}{m} \Bigl \{ \c^{\a} c_{\a}  \Bigl   [
[  m {\ov A}  + g  {\ov A}^2 ]
  [g'_1  m^2 +  g'_2  m {\ov A}
+  g'_3   {\ov A}^2 \Bigr ]
\]
\be
+ {g''}_1 m^3 {\ov A}    \}
\ee
Now the physically meaningfull quantity should be
${g''}_1$.  We find that it is given by:
\be
{g''}_1 = \fr{1}{g^3} [ g_4 - g g_3 + g^2 g_2 - g^3 g_1 ]
\la{g1}
\ee
Note the close relation between (\ref{g4})
and (\ref{g1}).  In particular, since $g$ is a
physical coupling constant, it is clear that gauge
invariance of one implies gauge invariance of the
other.   In this formulation the isomorphism
(\ref{iso}) is realized in a less obvious way.

It would be possible
to introduce more factors of $\fr{1}{m}$ in this
context and generate an infinite series, and
in this way `eliminate  the cohomology space'.
Clearly once this process starts it must be continued to
all powers of  $\fr{1}{m^k}$  to completely `eliminate
all anomalies'.
However this would  not be a natural procedure
in the present  context, because
no  more factors of $\fr{1}{m}$ should be introduced than
were present in the starting operator (\ref{opstart}).
Since the theory
is renormalizable so long as we do not propagate the
$\F_{\a}$ field and restrict ourselves to treating
it to first order, there is no justification for introducing
such an infinite series of renormalizations of arbitrary
non-renormalizable order.

Incidentally it seems unlikely that an operator as
simple as (\ref{opstart}) will be likely to develop
an anomaly, because the diagrams for it are not
linearly (or more) divergent, which is probably
necessary to develop an anomaly, if we can judge
from the known cases.

\section{The Space ${E}_{\infty \rm Special}$ }

It is not at all obvious, looking at the highly complicated operator
$\d$ above, whether one can find a subspace of the cohomology space
$H= {\rm ker} [\d + \d^{\dag}]
\approx E_{\infty}$
 without finding the solution to the whole problem.
But, fortunately, this can in fact be done here, as we now explain.

We shall concentrate
on the following kind of polynomial and shall find a set of restrictions on
it which imply that it will belong to $E_{\infty}$.
\be
{E}_{0 \; \rm Special}
= P[ A, \y, F, {\tilde A}, {\tilde \y}, {\tilde F},
   N({\ov c}) \geq  1, N(\e) = 4 ]
\ee
and we will find equations for the spaces
\be
{E}_{r \; \rm Special}
=
{E}_{0 \; \rm Special}
\cap
{E}_{r }
\ee
What we mean here is that this polynomial depends only on the field
variables
shown and no others, that it contains no derivatives $\pa_{\m}$,
that it has $ N({\ov c}) \geq 1$
and $ N(\e) = 4 $. Of course the  complex conjugate of the
above works the same way and we shall not treat it separately.

The problem we must confront is that these polynomials do get mixed
with others by the operators $d_r$ in general, and we have to
demonstrate that, with suitable restrictions, all the  operators
$d_r$ and $(d_r)^{\dag}$ do
in fact give zero on the subspace
${E}_{\infty \rm Special}$.
We will now find a further set of constraints which we can impose to
ensure that this is indeed a subspace of $E_{\infty}$.
\section{ The Space $E_{ 2 \rm Special} =
E_2 \cap {E}_{0 \rm Special} $}

Our next concern is with the last five
terms of (\ref{d1})
 since all the operators
$ c^{\a}	\L_{\a\;{\rm
Gauge}} + {\ov	c}^{\dot	\a}	{\ov	\L}_{\dot
\a\;{\rm Gauge}} + c^{\a}	{\tilde
\L}_{\a\;{\rm Gauge}} + {\ov	c}^{\dot	\a}
{\tilde { \ov	\L} }_{\dot	\a\;{\rm Gauge}}
$ (and their adjoints) certainly give zero
on our subspace
${E}_{\infty \rm Special}$
because it contains no gauge fields.
First we have the operator
\be
  \P_1 \e^{\m} \pa_{\m}  \P_1
\ee
This is zero on our subspace because it has
$N(\e)=4$ and the adjoint is zero because our subspace contains no
derivatives.
 The terms
\be
\int d^4 x \Bl [ {\ov F}_i
\fr{\d}{\d {\tilde A}_i}
+{  F}^i
\fr{\d}{\d {\tilde {\ov A}}^i}
 \Br ]
\ee
are zero in the antichiral subspace  that has  $N(c) \geq 1$
and in the  chiral subspace  that has $N{\ov c} \geq 1$, since it
takes chiral to antichiral or vice versa (same for adjoints).
Finally we have the terms:
\be
\P_1
 \int d^4 x \{
 m T^{ai} \w^a
\fr{\d}{\d A^i}
 + m {\ov T}^{a}_{i} \w^a
\fr{\d}{\d {\ov A}_i }
\Br \}
\P_1
\la{goldop}
\ee
These are the  well known homogeneous
terms that prevent the appearance of mass
terms for Goldstone bosons in a
spontaneously broken theory.   The situation here is
more complicated because we are already in
$E_1$ so that the projection operators have a non-trivial
effect in this operator.  We will simply impose the
equations
\[
[ T^{a}_{i}  A^{ i }]^{\dag}
{E}_{2 \rm Special}
=
[ T^{a}_{i}  \y_{\a}^{ i }]^{\dag}
{E}_{2 \rm Special}
=
[ T^{a}_{i}  { F}^{i} ]^{\dag}
{E}_{2  \rm Special}
\]
\be
=
[ T^{ai}   {\tilde A}_i ]^{\dag}
{E}_{2  \rm Special}
=
[ T^{ai}  {\tilde \y}_{i \a} ]^{\dag}
{E}_{2  \rm Special}
=
[ T^{ai}  {\tilde F}_{i} ]^{\dag}
{E}_{2  \rm Special}
=
0
\la{Teq}
\ee
which are stronger than we need.
These equations eliminate
from the cohomology space all
those chiral multiplets (and their sources) which contain the
Goldstone bosons of the spontaneous breaking
of the gauge symmetry.
  So these equations eliminate any dependence
on the Goldstone type multiplets,
but still allow dependence
of ${E}_{2  \rm Special}$
on all the other chiral multiplets.
A complete treatment of these equations here would lead us into
a treatment of the gauge fields too, since $\w$ plays an
important role in the operator above in
(\ref{goldop}).

\section{ The Space $E_{ 3 \rm Special} =
E_3 \cap {E}_{0 \; \rm Special}$}

The next term in the expansion of $\d$ is:
\be \d_2= \int d^4 x \Bl \{
\pa_{\m} Z^a
\fr{\d}{\d {\tilde V}_{\m}^a}
+ \Box \x^a
\fr{\d}{\d {\tilde \w}^a}
\Br \}
\ee
It is easy to see that
\be
\d_2
{E}_{2  \rm Special}
=
\d_2^{\dag}
{E}_{2  \rm Special}
=0
\ee
because this space contains no gauge fields or ghosts.
But the complete $d_2$ operator is actually of the form
\be
d_2 = \P_2 \{ \d_1 \fr{\d_0^{\dag}}{\D_0} \d_1 - \d_2 \} \P_2
\ee
Therefore, to check that
\be
d_2  {E}_{2\; \rm Special}
=
d_2^{\dag}  {E}_{2\; \rm Special}
=0
\ee
requires
a bit more work.  These do in fact give zero because the
other terms in $d_2$ all mix chiral with antichiral fields.

\section{ The Spaces $E_{ r\; \rm Special} =
E_r \cap {E}_{0 \rm Special}$}

For $r \geq 3$, the formula for
$d_r$ becomes increasingly complicated in terms of the
operators $\d_r$ and the Laplacians \ci{cmp1}.
We will imagine for present purposes that we can simply
take
\be
d_r \approx \P_r \d_r \P_r
\ee
and we will not need (at present) to return to the problem of
analyzing  the correct expression for $d_r$--the point is that we
will find a subspace of  $H$ in the following process, and we will
therefore be able to verify our result without the spectral sequence once we
have found it.  But the spectral sequence helps us to
organize the  task and enables us to use results that
have already been  derived.

The next operators are:
\be
\d_3= \int d^4 x \Bl \{
k  Z^a
\fr{\d}{\d {\tilde Z}^a}
+  \s^{\m}_{\a \dot \b} c^{\a} \x^a
\fr{\d}{\d {\tilde {\ov \l}}^a_{ \dot \b} }
\Br \}
\ee
\[ \d_5	=
c^{\a}  {\ov \na}_{\a \;{\rm Gauge}}
+ {\ov	c}^{\dot
\a}	{\na}^{\dot	\a }_{\;{\rm Gauge}} +
c^{\a}	{\tilde {\ov \na}}_{\a\;{\rm Gauge}}
+ {\ov c}^{\dot \a}
{\tilde  {
\na}}^{\dot \a }_{\;{\rm Gauge}}
\]
\be
+  c^{\a}
\s^{\m}_{\a \dot \b}
{\ov c}^{\dot \b} (\e^{\m})^{\dag}
\ee
Since ${E}_{r\; \rm Special}$ does not depend
on gauge or ghost fields, we easily see that
$ d_r \approx\P_r  \d_r \P_r ;  \;   r= 2,3,4,5$ and their adjoints
are all zero on our subspace ${E}_{r \;\rm Special}$.

Next we have
\[
\d_6 =
c^{\a}   {\ov \na}_{\a \;{\rm Matter}}
+ {\ov	c}^{\dot
\a}	{\na}^{\dot	\a }_{\;{\rm Matter}} +
c^{\a}	{\tilde {\ov \na}}_{\a\;{\rm Matter}}
+ {\ov c}^{\dot \a}
{\tilde  {
\na}}^{\dot \a }_{\;{\rm Matter}}
\]
\[
+ \int d^4 x
\Bl \{
 c^{\a}
\s^{\m}_{\a \dot \b}
{\ov c}^{\dot \b}   {\ti \w}^a
\fr{\d}{\d {\ti V}^{\m a} }
\]
\[
+   c^{\a}
\s^{\m}_{\a \dot \b}
{\ov c}^{\dot \b}
 {  V}_{\m}^a
\fr{\d}{\d { \w}^a}
+   c^{\a}
\s^{\m}_{\a \dot \b}
{\ov c}^{\dot \b}
 \pa_{\m} {\ti  Z}^a
\fr{\d}{\d {\ti \x}^a}
\]
\be
+   c^{\a}
\s^{\m}_{\a \dot \b}
{\ov c}^{\dot \b}
 \pa_{\m} {\x}^a
\fr{\d}{\d {Z}^a}
\Br \}
\ee
The first part of this operator was discussed in \ci{dmr}.  It is
\[
d_{6  ;{\na} }= \P_6 \d_{6  ;{\na} } \P_6 =
\P_6  \{ c^{\a}   {\ov \na}_{\a \;{\rm Matter}}
+ {\ov	c}^{\dot
\a}	{\na}^{\dot	\a }_{\;{\rm Matter}}
\]
\be
+
c^{\a}	{\tilde {\ov \na}}_{\a\;{\rm Matter}}
+ {\ov c}^{\dot \a}
{\tilde  {
\na}}^{\dot \a }_{\;{\rm Matter}}
\} \P_6 \ee
The Laplacian for this part is, when operating on chiral polynomials
that contain no derivatives:
\[
\D_{6; \na}	=
\P_6
\Bl	\{
({\na}_{\dot	\a})^{\dag}	\P_6	{\na}_{\dot	\b}
+
{\ov	n}	[	M	-4	]
\]
\be
+
({\ov	\na}_{\a})^{\dag}	\P_6	{\ov	\na}_{\a}
+
{n}	[	{\ov	M}	-4	]
\Br	\}
\P_6	.	\label{del6}
\ee
where we 	use	the	following	abbreviations
\be
M	=	 	4	N(F)	+	2	N(\y)
\ee
\be
{\ov	M}	=		4	N({\ov	F})
+	2	N({\ov	\y})
\ee
As is discussed at length in \ci{dmr},
this operator restricts us to functions that have at most one
$F$ or two $\y$ in them ( but still any number of $A$ fields).
Also there is a nontrivial condition from the equation
\be
\P_6
({\na}_{\dot	\a})^{\dag}	\P_6
{E}_{ 7 \; \rm Special}=0
\ee
This equation and all the others so far can be solved by
taking a product of those
chiral superfields  which are not Goldstone superfields of the
spontaneous gauge symmetry breaking,
with no derivative operators, and then
integrating the result over chiral superspace.  The result will
be in
${E}_{7 \rm Special}$.
The rest of $d_6$ and its adjoint clearly give zero on our
special subspace.

The next operator yields a number of important restrictions.
It is:
\[ \d_7 = \int d^4 x \; \Bl \{
c^{\a}
D^a
\fr{\d}{\d {\ti D}^a}
+
\s^{\m}_{\a \dot \b}
 \pa_{\m} {\ov \y}^{\dot \b}_i
\fr{\d}{\d {\ti \y}_i}
\]
\[+ m {\ov g}^{ij} {\ov  F}_j
\fr{\d}{\d  {\ti {\ov  F} }^i}
+ m {\ov g}^{ij} { \ov A}_j
\fr{\d}{\d  {\ti   {\ov  A} }^i}
 + m {\ov g}^{ij} { \ov \y}^{\dot \a}_j
\fr{\d}{\d  {\ti {\ov  \y}}^{ i\dot \a } }
\]
\[
  + m { g}_{ij} F^j
\fr{\d}{\d  {\ti  F }_i}
 + m {  g}_{ij} { A}^j
\fr{\d}{\d  {\ti   { A} }_i}
 + m {  g}_{ij} { \y}_{ \a}^j
\fr{\d}{\d  {\ti { \y}}_{ i \a } }
\]
\be
+ c^{\a}  	T^{ai}	\s^{\m}_{\a	\dot{\b}}	{\tilde {\ov
\y}}_{i	}^{\dot	\b}
	\fr{\d	}{\d	{\tilde	V}^{a	\m}	}
\Br \}
\ee
This is very similar to the problem that we analyzed
above in section (\ref{masssec}) except that there are
more fields and the mass matrices
must be diagonalized.  For simplicity of
notation, let us assume that it is diagonal.
Let us define the operator $\W$ by:
\[
m \W =
\int d^4 x \{ m { g}_{ij} F^j
\fr{\d}{\d  {\ti  F }_i}
 + m {  g}_{ij} { A}^j
\fr{\d}{\d  {\ti   { A} }_i}
 + m {  g}_{ij} { \y}_{ \a}^j
\fr{\d}{\d  {\ti { \y}}_{ i \a } }
\}
\]
\be
\approx
m \sum_j
e_j  [
F^j
{\ti  F }_j^{\dag}
+
A^j
{\ti  A }_j^{\dag}
+
\y^{j \a}
{\ti \y }^{ \a \dag}_j
]
\ee
Note that this operator commutes with
supersymmetry
\be
\{\L_{\a}, \W\} = 0
\ee
so that it is easy to find the solution at this point.
The relevant form is
\[
{E}_{8 \rm Special}
= P_{\rm Massless} [m,  A, \y, F, {\tilde A}, {\tilde \y}, {\tilde F},
   N({\ov c}) \geq  1, N(\e) = 4 ]
\]
\be
+\W P_{\rm Massive} [ A, \y, F, {\tilde A}, {\tilde \y}, {\tilde F},
   N({\ov c}) \geq  1, N(\e) = 4 ]
\la{e8special}
\ee
where by
$ P_{\rm Massless} $ is we mean that this arbitrary polynomial
can depend only on massless chiral superfields but also on
the parameter $m$, whereas by
$ P_{\rm Massive} $ we mean that if any monomial
in this arbitrary polynomial
depends on a massive chiral superfield, then
it cannot also depend on the parameter $m$.
This follows from the following form
of the Laplacian:
\be
\D_7 = \P_7 \{
[N_{\rm Massive}(A) + N_{\rm Massive}({\tilde A})] N(m)
+ \W^{\dag} \W
\}
\P_7
\ee
where
\be
N_{\rm Massive}(A) + N_{\rm Massive}({\tilde A} )
= [ \W + \W^{\dag}]^2
\ee
So at this stage we have a cohomology space which includes
both massive fields to be treated in the rather complicated way indicated
in section
(\ref{masssec}) and also massless fields which can occur
accompanied by explicit powers of mass and
consequently may be easier to separate from other
terms.  The equation of motion terms in $\d_7$
again are automatically zero in the subspace ${E}_{7 \rm Special}$
because they
mix chiral with antichiral fields or sources.
Next we have
\[ \d_9 =  \int d^4 x \; m
\Bl \{ 	{\ov	T}^{a	}_i	\l^a_{\a}
	\fr{\d	}{\d	{\tilde	\y}_{i	\a}	}
+ { 	T}^{a	i}	{\ov \l}^a_{\dot \a}
	\fr{\d	}{\d	{\tilde	{\ov \y}}^i_{\dot \a}	}
\]
\be
+ {\ov	T}^{b	}_i	\y^{j}_{\a}
	\fr{\d	}{\d	{\tilde	\l}^a_{	\a}	} +	T^{ai}
{\ov	\y}_{i	\dot	\a}
	\fr{\d	} {\d
{\tilde	{\ov	\l}}^a_{	\dot	\a} }
\Br \}
\ee
\be
 \d_{10} = \int d^4 x \;  m
 \Bl \{	D^a	{\ov T}^{a	}_i
	\fr{\d	}{\d	{\tilde	F}_i	} +{\rm c.c.}
+	(	{\ov	T}^{a	}_i	A^i	+	T^{ai}
	{\ov	A}_i	) \fr{\d	}{\d	{\tilde	{D}}^a	}
 \Bigr	\}
\ee
These are eliminated because of (\ref{Teq}).
\be
\d_{11} = \int d^4 x \;
\Bl \{  \s^{\m}_{\a	\dot{\b}}	\pa_{\m}	{\ov	\l}^{a
\dot	\b}
	\fr{\d	}{\d	{\tilde	\l}^a_{	\a}	} +	{\ov
\s}^{\m}_{\dot	\a	{\b}}	\pa_{\m}	{	\l}^{	a
\b}
	\fr{\d	}{\d	{\tilde	{\ov	\l}}^a_{	\dot	\a}	}
\Br \}
\ee
This is eliminated because our special subspace contains
no gauge fields.
\be
\d_{13} = \int d^4 x \;
	\{
\Box	{\ov	A}_i
	\fr{\d	}{\d	{\tilde	F}_i	}
	+ \Box	{ 	A}^i
	\fr{\d	}{\d	{\tilde	{\ov F}}^i	}
\}
\ee
These equation of motion terms
are zero in the subspace because they
mix chiral with antichiral fields.
\be
\d_{14} = \int d^4 x
\Bl \{
\pa_{\m}	V^{\m	a}
	 	{\ov T}^{a	}_i
	\fr{\d	}{\d	{\tilde	F}_i	} + {\rm c.c.}
+	m [T^{ai}	\pa^{\m}	{\ov	A}_i
	+	  {\ov	T}^{a	}_i	\pa^{\m}	{A}^i	]
	\fr{\d	}{\d	{\tilde	V}^{\m	a}	}
\Br \}
\ee
Again, this is zero because of (\ref{Teq}) or because
it contains derivatives.
\[
\d_{15} = \int d^4 x \;
\Bl \{		(	\Box	V^{a	\m}	-	\pa^{\m}	\pa^{\n}
V^a_{\n})
	\fr{\d	}{\d	{\tilde	V}^{\m	a}	}
\]
\be
+ k	m {\ov	T}^{b	}_i	c^{\a} \y^{i	\a} \fr{\d	}{\d
{\tilde	\x}^{a}	} + {\rm c.c.}
+ 	k	m (	T^{ai}	{\ov	A}_i +	{\ov	T}^{a	}_i	A^i	)
	\fr{\d	}{\d	{\tilde	Z}^a	}
\Br \}
\ee
Again, this is zero because of (\ref{Teq}).
\be
\d_{16} = \int d^4 x \;
\Bl \{	 	\pa^{\m}	V^a_{\m}
 \fr{\d}{\d	{\tilde	Z}^a	}
+	\Box	\w^a	\fr{\d}{\d	{\tilde	\x}^a	}
\Br \}
\ee
This is zero because it contains none
of the fields in our special subspace.
Now we come to an important operator:
\[
\d_{17} = \int d^4 x \;
\Bl \{
-	[ f^{abc}	{\tilde	\lambda}^{b}_{\a}	\l^{\a
c} +	f^{abc}	{\tilde	{\ov	\l	}	}^{b	\dot	\a}
	\ov{\lambda}^{c}_{\dot	\a} ]
\fr{\d}{\d	{\tilde	\w}_{ a} }
\]
\[	-	\frac{1}{2}	f^{abc}
\omega^{b}	\omega^{c}
 \fr{\d}{\d	\w^a }
+  \w^a J^a
+ c^{\a} \s^{\m}_{\a \dot \b} \pa_{\m} {\ov \l}^{a \dot \b}
\fr{\d}{\d	{\tilde	\x}_{ a} }  + {\rm c.c.}
+ c^{\a}
\s^{\m}_{\a \dot \b} T^{ai}_j V^a_{\m} { \ov  \y}_i^{ \dot   \b}
\fr{\d}{\d	{\tilde	F}_{ j} }
\]
\[ + k	m {\ov	T}^{b	}_i	\y^{i	\a} \fr{\d	}{\d
{\tilde	\x}^{a}	}  + {\rm c.c.}
\]
\[
+ \Bl [
 f^{abc}
{\tilde	D}^{b} D^{c} + V^{\m	a	b}	 {\tilde
V}_{\m}^b
+	{\tilde	\y}_i^{\a}
T^{ai}_j	\y_{\a}^j
+	{\tilde	{\ov \y}}^{i \dot \a}
{\ov T}^{aj}_i {\ov \y}_{j \dot \a}
\]
\be
	 + {\tilde	A}_i
	T^{ai}_j	F^j
	+ {\tilde	{\ov A}}^i
	{\ov T}^{aj}_i {\ov F}_j
	+   A^i
	{  T}^{aj}_i {\ti  F}_j
	+ {\ov	A}_i
	{\ov T}^{ai}_j	{\ti {\ov F}}^j
\Br ]
 \fr{\d}{\d	{\tilde	\w}^a }
\Br \}
\ee
where
\be J^a(x)  =
 \sum_{ {\rm All\;\;Fields\;(except }
\; \x, Z, {\ti \x} , {\ti Z} {\rm )} }
 {\rm (Field)}^j(x)	T^{a i}_j \omega^{c}
 \fr{\d}{\d	{\rm (Field)}^i(x) }
\ee
Two new restrictions arise from this operator.
They are:
\be
J^a
{E}_{18 \rm Special}=0 \;\;({\rm a=non-goldstone\; directions\;
only} )
\la{rotate}
\ee
At this point it is necessary to point out that
only those $\w^a$ which are still gauge symmetries
survive to this stage.  The others were eliminated when the
Goldstone modes were eliminated, when the equations
(\ref{Teq}) were imposed.
  Next we have: \[
\d_{18} =
\int d^4 x \;
\Bl  \{
  \;	k	\;	m
\;	[ \xi^{b}
	T^{bj}_i	u^i T^{a	k}_j	 	{\ov	A}_k  +
\xi^{b}	{\ov	T}^{bj}_i	{\ov	u}_j
T^{ai}_j	 	A^j   ]
 \fr{\d}{\d	{\tilde	\w}^a }
\]
\be
 +	k	\;	m	\;	 \xi^{a}	{\ov	T}^{a}_j
T^{bj}_i	\w^b
 \fr{\d}{\d	{\tilde	F}_i }
	+	\frac{1}{2}	f^{abc}
V^b_{\mu} V^c_{\nu} 	\s^{\mu	\nu}_{\a	\b}
c^{\b}  \fr{\d}{\d \lambda^{a}_{\a} }
\Br \}
\ee
\be
\d_{19} =
\int d^4 x
 V^{\m	a	b}	 \pa_{\m} \x^b
\fr{\d}{\d	{\tilde	\w}^a }
\ee
These both yield zero.
\[
\d_{22} =
\int d^4 x \;
	\Bl \{
   	f^{abc} {\tilde
D}^{b}	[ \fr{-i}{2}	c^{\a}	\s^{\mu}_{\a
\dot{\b}	} \ov{\l}^{c	\dot{\b}	} +\fr{i}{2}
\l^{c	\a	} \s^{\mu}_{\a	\dot{\b}	}
\ov{c}^{\dot	\b}	]
\fr{\d}{\d	{\tilde	V}_{\m}^a }
\]
\be
 +(V_{\m}	{\tilde	D})^{	a	} \s^{\mu
}_{\a	\dot	\b}	{\ov	c}^{\dot	\b}
\fr{\d}{\d
		{\tilde	\lambda}^{a}_{\a} }
 -
\Bl [
\fr{-i}{2}	c^{\a}	\s^{\mu}_{\a
\dot{\b}	} V_{\m}^{ab}	\ov{\l}^{b	\dot{\b}	}
+\fr{i}{2}	V_{\m}^{ab}	\l^{b	\a	}
\s^{\mu}_{\a	\dot{\b}	} \ov{c}^{\dot	\b}
\Br ]
 \fr{\d}{\d	  D^{a} }
\Br \}+ {\rm c.c.}
\ee
This yields no restriction. Next
\[
\d_{23} =
\int d^4 x \;
\Bl \{
\Bl [
V^j_{\mu	i} {\ov	V}_j^{\m	k	}
 	  m	{\ov	u}_k
+ g_{ijk} [2  A^j F^k +
\y^{\a	j}	\y_{\a}^k ]
\Br ]
\fr{\d}{\d	  {\tilde	F}_i	 }
+
	(V_{\m}	{\tilde	A})_i \s^{\m}_{\a	\dot{\b}}
\overline{c}^{\dot{\b}}
\fr{\d}{\d	  {\tilde	\y}_{i \a}	 }
\]
\be
+
  g_{ijk}
A^j	\y_{\a}^k
 \fr{\d}{\d	{\tilde	\y}_{i \a} }
+
  (	V_{\m}	\y)^{i	\a}
\s^{\m}_{\a	\dot{\b}}
\overline{c}^{\dot{\b}}
\fr{\d}{\d	  F^i	 }
+
  g_{ijk} 	A^j	A^k
\fr{\d}{\d	 {\tilde	A}_i	 } + {\rm c.c.}
\ee
This may generate a non-trivial constraint since
the argument made in section (\ref{masssec})
will not work in general.  We shall not attempt
a general treatment of it here, except to mention
that again this operator commutes with supersymmetry.
\be
\{ \d'_{23} ,  \L_{\a} + {\tilde \L}_{\a} \} =0
\ee
where we define
\[
\d'_{23} =
\int d^4 x \;
  g_{ijk}
\Bigl \{
A^j	\y_{\a}^k
 \fr{\d}{\d	{\tilde	\y}_{i \a} }
\]
\be
+
    \Bigl [2	A^j	F^k
+ \y^{\a	j}	\y_{\a}^k \Bigl ]
\fr{\d}{\d	  {\tilde	F}_i	 }
+
 A^j	A^k
\fr{\d}{\d	 {\tilde	A}_i	 } + {\rm c.c.}
\Bigr \}
\P_{23}
\la{del23}
\ee
Next
\be
\d_{24} =
\int d^4 x \;
\Bl \{
 (V_{\m}
A)^i \s^{\m}_{\a	\dot{\b}}
\overline{c}^{\dot{\b}}
\fr{\d}{\d	  \y^i_{\a}		 }
+
\sigma^{\m}_{\;\a	\dot	\b}	(V_{\mu}
\ov{\y}^{\dot	\b})_i
\fr{\d}{\d	  {\ti \y}_{i \a}		 }
\Br \} + {\rm c.c.}
\ee
The first term needs some thought.  It is of course required
to make gauge invariant the transformation of the $\y$ field
when derivatives are present.  For our special subspace
with no derivatives, it can be ignored.  The reason is
that for $k =1$, (\ref{vequation}) eliminates all $V_{\m}$ with
no derivatives from $E_1$ (See \ci{cmp1} for details).
The second term mixes chiral and antichiral
and so it can be ignored as usual.
Now, for reasons already given, all terms vanish up to $\d_{39}$.
\be
\d_{25} =
\int d^4 x \;
\Bl \{
{\ov
\y}^{\dot	\a}_{	j}	T^{aj}_{i}	{\ov
\l}^a_{\dot	\a	}
\fr{\d}{\d	  {\tilde	F}_i	 }
	+
 {\ov
T}^{ai}_{j}	 	{\ov	A}_i	 	{\tilde	{\ov	\y}
}^{i	\dot	\a}	{\ov	\s}_{\m	\dot	\a	\b}
	c ^{	\b	}
\fr{\d}{\d	  {\ti V}^a_{\m}	 }
+
	T^{a	i}_j	 {\ov 	A}_i 	\y^{\a	j}
\fr{\d}{\d	  {\tilde	\lambda}^{a}_{\a} }
\Br \} + {\rm c.c.}
\ee
\be
\d_{26} =
\int d^4 x \;
\Bl \{
 D^a	T^{aj}_{i}
 	{\ov	A}_j
\fr{\d}{\d	  {\tilde	F}_i	 }
+
	T^{ai}_{j} \y^{\a	j}
\sigma^{\m}_{\;\a	\dot	\b}	\ov{\y}^{\dot
\b}_i
\fr{\d}{\d	  {\ti V}^a_{\m}	 }
+
 	T^{aj}_{i}
 {\ov	A}_j
 	A^i
\fr{\d}{\d	 {\tilde	D}^{a}	 	 }
\Br \} + {\rm c.c.}
\ee
\be
\d_{28} =
\int d^4 x \;
\Bl \{
 -	\frac{1}{2}	f^{abc}
\lambda^{b	\a} \s^{\m}_{\a	\dot{\b}	}
\ov{\lambda}^{b	\dot{\b}	}
\fr{\d}{\d	  {\ti V}^a_{\m}	 }
+
\s^{\m}_{\a	\dot{\b}	}	(V_{\m}	{\ov	\l})^{a
\dot	\b}
\fr{\d}{\d	   {\tilde	\lambda}^{a}_{\a}}
\Br \}
\ee
\be
\d_{30} =
\int d^4 x \;
\Bl \{
 [ V^j_{\mu	i}	{\ov	\pa}_j^{\m	k	}
 	{\ov	A}_k  +\pa_{\mu	 }	{\ov	V}_i^{\m	k	}
 	  {\ov	A}_k  ]
 \fr{\d}{\d	  {\ti F}_i	 }
 +	[T^{ai}_{j}	 A^j 	\pa^{\m}    {\ov	A}_i +
{\ov T}^{ai}_{j}	 {\ov	A}_i \pa^{\m}
A^j 	  ]
\fr{\d}{\d	  {\ti V}^a_{\m}	 }
\Br \}
\ee
\[
\d_{32} =
\int d^4 x \;
	 	T^{ai}_{j}	\l^a_{\a}
 	{\ov	A}_i
\fr{\d}{\d	   {\tilde	\y}_{i \a}}
+ \Bl [  f^{abc} V^{b \n}	[
\pa_{\mu}	V_{\nu})^{c} - \pa_{\nu}
V_{\mu})^{c} ]  + f^{abc} \pa^{\nu} [
V^{b}_{ \m}  	V_{\nu})^{c} \Br ]
\fr{\d}{\d	   {\tilde	V}_{\mu}^{a} }
\]
\be +
\Bl [
k\;	m\;
	T^{ai}_{j}	u^j {\ov	T}^{ak}_{i}	\w^a
 	{\ov	A}_k	   +	k\;	m\;	{\ov
T}^{ai}_{j}	{\ov	u}_i
	T^{aj}_{k}	\w^a {	A}^k
\Br ]
\fr{\d}{\d	  {\tilde	\x}^a}
\Br \}
\ee
\be
\d_{33} =
\int d^4 x
\pa_{\m}		V_{\mu}^{ab}	\omega^{b}
\fr{\d}{\d	  {\tilde	\x}^a}
\ee
\be
\d_{35} =
\int d^4 x \;
\Bl [   (V_{\n}	{\tilde
\lambda})^{	a	\a}
	\s^{\mu	\nu}_{\a	\b}	c^{\b} +	(V_{\n}
{\tilde	{\ov	\l	}	})^{a	\dot	\a}
	\ov{\s}^{\mu	\nu}_{\dot{\a}	\dot{\b}	}
\ov{c}^{\dot	\b} \Br ]
\fr{\d}{\d	   {\tilde	V}_{\mu}^{a} }
\ee
All the above give no new constraints.  However $\d_{39}$ does
yield some new equations.
\[
\d_{39} =
\int d^4 x \;
\fr{12}{m} g_{ijkl}
\Bigl \{
(		A^j A^k	F^l +
\y^{\a	j}	\y_{\a}^k A^l	)
\fr{\d}{\d	   {\tilde	F}_i }
\]
\be
+
	A^j A^k
\y_{\a}^l
\fr{\d}{\d	   {\tilde	\y}_{\a}^k }
+
\fr{1}{3} 	A^j A^k
A^l
\fr{\d}{\d	   {\tilde	A}_i}
\Bigr \}
\la{del39}
\ee
and once again this operator commutes with supersymmetry.
Finally we have
\be
\d_{49} =
\int d^4 x \;
 f^{abc}  V^{b \n} f^{cde}  V^{d}_{\mu}
V_{\nu})^{e}
\fr{\d}{\d	   {\tilde	V}_{\mu}^{a} }
\ee
\be
\d_{50} =
\int d^4 x \;
\Bl [	T^{ai}_{j}	 	A^j 	{\ov
V}_i^{\m	k	}
 	{\ov	A}_k  +	{
	V}_i^{\m	k	}	 	A^i 	{\ov
T}^{aj}_{k}
 	{\ov	A}_j
\Br ]
\fr{\d}{\d	   {\tilde	V}_{\mu}^{a} }
\ee
and these give nothing new.

\section{Equations for
${E}_{\infty \rm Special}$ }

Collected together we get:

\be
\L_{\a \rm Matter} {E}_{\infty \rm Special}	=	0
\ee
\be
\P_6	\na_{\dot	\b {\rm Matter} }
{E}_{\infty \rm Special}=0
\ee
\be
\P_6 [ M -4 ]{E}_{\infty \rm Special}=0
\ee
where we 	use	the	following	abbreviation
\be
M	=	 	4	N(F)	+	2	N(\y)
\ee
So this restricts us to functions that have at most one
$F$ or two $\y$ in them.
However they can still have any number of
$A$ fields.
Next, we also have the equations that
result from the requirement that this
subspace be independent of the  Goldstone modes
and not vanish by the equations of
motion. The relevant subspace was (\ref{e8special})
as restricted by equations (\ref{Teq})
and (\ref{rotate}),
but I have not tried to further explicitly restrict this
with the operators (\ref{del23}) and
(\ref{del39}).  That is best left for
specific cases where the tensors are known
explicitly.

\section{The Space $H_{\rm Special}$ }
\la{hspecial}

Now we give rules to construct the space
$ H_{\rm Special} \approx
 E_{\infty \rm Special}$.
The simplest case to analyze is the case when there is no chiral action
at all and consequently no spontaneous gauge symmetry breaking or
VEVs. In this case the answer is already known from previous work,
and the answer is that $H_{\rm Special}$
contains
all possible expressions of the form
\be \int	d^4	x	d^2	{\ov	\q}	\F^{\a}
{\ov	S}_{i_1} \cdots	{\ov	S}_{i_1}
t^{ i_1,  i_2,  	\cdots	i_k } c_{\a} \ee
where the tensors
$t^{ i_1,  i_2,  	\cdots	i_k }$
are invariant
tensors under the group:
\be
\sum_{q=1}^k
T^{a i_q}_{\;\;\;\; j}
t^{ i_1,  i_2, \cdots j {\hat i_q}	\cdots	i_k }
=0
\la{invarten}
\ee

What we have discovered by the foregoing analysis is very
simple. When $S_{\rm Chiral}$ is non-zero,
the set of constraints on
the cohomology space  $H_{\rm Special}$
changes in the following ways:
\begin{enumerate}
\item
When $S_{\rm Chiral}$ is non-zero,
one gets the additional constraints
(we assume the renormalizable case again here):
\be
[m g_{ij}(u) F^{j }
+ g_{ij k}(u)   ( A^j F^k + \y_{\a}^j \y^{ k \a}) ]^{\dag}
H_{\rm Special} =0
\ee
\be
[ m g_{il}(u)  A^l + g_{ij k}(u)     A^j A^k  ]^{\dag}
H_{\rm Special} =0
\la{aeq}
\ee
\be
[ g_{il}(u)    \y_{\a}^l +   g_{ij k}(u)     A^j
\y^k_{\a}  ]^{\dag} H_{\rm Special} =0
\ee

  In general I would expect that there are
plenty of solutions-- equivalently there are usually
plenty of supersymmetric  polynomials with the required
gauge invariance that are not related
by the equations of
motion.    The simplest such objects will
be those which commence with a zero mass superfield.
\item
\la{gold}
When the VEV is not zero, one gets the additional
constraints
\be
[T^{a}_{i} A^i]^{\dag}
H_{\rm Special} =0
\ee
\be
[T^{a}_{i} \y^{i \a}]^{\dag}
H_{\rm Special} =0
\ee
\be
[T^{a}_{i} F^i]^{\dag}
H_{\rm Special} =0
\ee
\item
Once the equations for (\ref{gold})
are satisfied, the invariance of the tensor in
(\ref{invarten}) reduces to invariance under the
remaining gauge invariance after gauge symmetry breaking
as required by equation (\ref{rotate}).
\item  It is clear that we do not generate the whole
cohomology space in this way, but only a part of it.
The rest must wait for a more complete solution of the
cohomology problem.
\end{enumerate}

\section{Discussion of the Superpotential}
\la{suppot}
Now that we have the equations for $H_{\rm Special}$,
and we have discussed the solutions,
let us discuss the superpotential from this point of view.
In most cases of interest, the reducible representation
$T^{ai}_{\;\;\;j}$ is fully reducible and reduces to a
set of irreducible representations of the group:
\be
S^i = \sum_I S^i_I
\ee
These irreducible chiral superfields then can have the following
properties:
\begin{enumerate}
\item
The first question for each $S^i_I$ is whether any of
its
$A_I^i$ components develop a VEV or not.  If
\be
<A_I^i> = m u^i \neq 0
\ee
then we shall call it a `Higgs' multiplet.
\item
The second question for each $S^i_I$ is whether any of
 its
components $A_I^i$ contributes to a Goldstone Boson or not.
If one or more of them do, we
shall call the multiplet a `Goldstone' multiplet.
Goldstone bosons always occur in Higgs multiplets because
that is the way that the Yang-Mills vector scalar mixing terms arise
in the shifted action before `t Hooft style gauge fixing.
\item
The third question for each $S^i_I$ is whether any
of its $A^i_I$ components do not appear in mass terms
after spontaneous gauge symmetry breaking even
though they are not Goldstone boson--these massless
but non-Goldstone bosons can be distinguished
by the fact that they do not have mixing terms with the
Yang-Mills vector fields, or equivalently, that the
multiplet possesses massless scalars but has zero VEV:
\be
<A_I^i> = o
\ee
Multiplets of this kind we shall call
massless matter chiral superfields.  Our
example below illustrates this phenomenon,
which is well known in the standard model
where there is no right handed partner for
the neutrino, so that it must be massless
even after spontaneous gauge symmetry
breaking.
\item
Finally there are multiplets which are none of the above.
These have zero VEV, no Goldstone bosons and no other
massless bosons.  These are the massive matter multiplets.
It is currently believed that the quarks are in massive
matter multiplets.
\end{enumerate}
In all the cases above, each boson has supersymmetry partners
of course, because supersymmetry is not broken.

Now return to our discussion of $H_{\rm Special}$.
Clearly the simplest solutions of these equations
involve massless  matter chiral
superfields, but there are also solutions that
involve massive matter multiplets in the
way indicated in section (\ref{masssec}).
We shall not try to discuss the higher
constraints  in any more detailed
way--one simply has to  verify that the equations are
solved for any particular case.

\section{A Simple Example}
\la{secexample}
This example was also presented in \ci{tamu45}.  The present
discussion will be more complete since the earlier discussion was
necessarily rather short by reason of space restrictions.

We consider a supersymmetric gauge theory
based on the gauge group $SU(2)$ with matter
in two vector multiplets and a singlet:
 $L^a : I = 1; H^a : I = 1; R : I = 0. $
   These `a' indices
transform with $i\e^{abc}$  and take the
values a=1,2,3.  Since the `a' indices are
real and since $\d_{ab}$ is an invariant tensor
of $SU(2)$, there is no difference when we raise and lower
these indices.

Without any good reason, we shall assume that the
superpotential
does not contain a mass term for the $L$ field.
It is easy to find more natural examples where
no mass term is possible for the relevant fields \ci{tamu47}.

Since the `Higgs field' $H^a$ is in a real
representation of the gauge group,
it can have a mass term in the
superpotential.

Now we assume the following form for
the superpotential:
 \be W = g_1  L^a H^a R +
\fr{g_2 m}{2} H^a H^a + \fr{g_3}{4 m} [H^a
H^a ]^2
\ee
Note that renormalizability is not a property of this superpotential.

If $g_2 g_3 <0 $ , the Higgs field will
develop a VEV in its  `A' term that breaks
the gauge symmetry down to $U(1)$ while
leaving  supersymmetry unbroken.  The $L$
and $R$ fields develop no VEV.
Let us denote components as follows:
\be
L^a = A^a + \q^{\a}  \y^a_{\a} + \fr{1}{2} \q^2 F^a
\ee
\be
H^a = B^a + m u^a + \q^{\a}  \f^a_{\a} + \fr{1}{2} \q^2 G^a
\ee
\be
R  = A + \q^{\a}  \y_{\a} + \fr{1}{2} \q^2 F
\ee
We distinguish $ a=i,3$ where $i=1,2$.
Here we have included a shift by
the  VEV:
\be < B^a>_{\rm before\; shift} =
\d^{a 3} m \sqrt {\fr{- g_2} {g_3} }
\equiv
\d^{a 3} m h
\equiv
m u^a
\ee

Then the `F' term of the superpotential becomes:
 \[ W_F = \Bigl [ g_1  L^a (H^a + \d^{a 3} m h )  R +
\fr{g_2 m}{2}
(H^a + \d^{a 3} m h )(H^a +  \d^{a 3} m h )
\]
\be
+ \fr{g_3}{4 m}
[(H^a +  \d^{a 3} m h )
(H^a +  \d^{a 3} m h )]^2
\Bigr ]_F
\ee
In terms of components, this makes the following contribution
to the action:
\[
S_{{\rm Chiral}}
=
\int d^4 x W_F =
\int d^4 x \Bigl \{ g_1
( m h A^3 F +
 m h \y^{3 \a} \y_{\a}
+  m h F^3 A )
\]
\[
+ g_1 (A^a B^a F
+ \y^a \f^a A
+ F^a B^a A
+ A^a \f^a \y
+ A^a G^a A
+ \y^a B^a \y  )
\]
\be
+ {\rm terms\; involving\; H \; superfield \; only}
\Bigr \}
\ee
The essential point to note here is that there
is no mass term like
\be
m (  A^i F +
  \y^{i \a} \y_{\a}
+  F^i A )
\ee
for $i = 1,2$
which would give a mass to the $L^1$ and $L^2$ superfields.
They are massless after spontaneous breaking of the gauge symmetry,
but they are not Goldstone fields.  The Goldstone bosons are
contained in the fields $H^i$.
Equally important for the example below is the fact that $L$
does not appear  quadratically in the superpotential so that
there is no term like $A^i A^i$ in the equations of motion
either.

The simplest composite operator (together
with a source $\F_{\a}$  that could develop
a supersymmetry  anomaly seems to be:
 \be S_{\rm	Composite}
	= \fr{1}{m^3}	\int	d^{4}x	d^4	\q \Bigl	\{	\F^{\a}
{\ov D}^{\dot	\b} 	[ 		{\ov	L}^a
	{\ov	H}^{a}  ]
{\ov	D}_{\dot	\b}
	D_{\a}	 R	\Bigr	\}
\ee
The fraction $\fr{1}{m^3}$ is included so that
$\F_{\a}$ will have its canonical dimension of $\fr{1}{2}$.
After translation of the Higgs field, we
find the terms \[
 S_{\rm	Composite}
	=	\fr{1}{m^2}  h \int	d^{4}x
\Bigl	\{
\c^{\a}
\Bigl	[
(\s^{\m
})^{\g	\dot	\b} \pa_{\m} {\ov A}^{ 3}
\s^{\n}_{\a	\dot	\b}
\pa_{\n} { \y}_{\g }
\]
\be
+
{\ov \y}^{3 \dot \b}
(\s^{\m })_{\a	\dot	\b} \pa_{\m} F
+ \cdots
\Bigr	] + \cdots
\Bigr	\}
\ee

The form of the supersymmetry anomaly that we would like to
calculate is
\be
\d \G_{\F}
	=	m e \int	d^{4}x	 d^2 {\ov \q}
\F^{\a}	 c_{\a}
\sum_{i = 1,2}
{\ov L}^{i } {\ov L}^{i }=
	m e  \int	d^{4}x
\c^{\a}	 c_{\a}
\sum_{i = 1,2}
{\ov A}^{i } {\ov A}^{i }
+ \cdots
\ee
This is in the cohomology space for the reasons given above--it
 is linearly independent of the polynomials that vanish
by the equations of motion of this theory, and
because these fields $A^i$ are massless,
there is no complicated mixing problem to separate their
coefficients like we need to do for massive fields in
section (\ref{masssec}).

The relevant triangle diagrams here are linearly divergent
and it appears dimensionally possible for the anomaly to
appear.  Do these diagrams add together to preserve
supersymmetry or not?  Should one introduce a regularization
somehow into the calculation? At present I am not sure
how to calculate the coefficient e in this or any model.
The first question clearly is `Which diagrams should
one calculate?'  This question will be the topic of
the next sections.

\section{Tactics for Computing Supersymmetry Anomalies}
\la{sfull}
Now let us consider how one would choose the diagrams
to actually compute a supersymmetry anomaly.
For this purpose, it is useful to simplify our operator
down to a more tractable size, to see how things work.

The following transformations are the gauge
and supersymmetry  transformations  for a very
simple  theory, involving only one chiral superfield.
We suppress the Yang-Mills fields and the Goldstone
mechanism for present purposes.
This operator is   the full BRS operator
which contains the partial operator analyzed in section
(\ref{masssec}).
\begin{equation}	\delta	A 	=	c^{\a}
\y_{\a}
\la{mew1}
\end{equation}
 \begin{equation} \delta	\y_{\a}	=
  \pa_{\m}	  A  \s^{\m}_{\a	\dot{\b}}
\overline{c}^{\dot{\b}} +	F	c_{\a}
\ee
\be \delta	F	= \pa_{\m}	\y^{  	\a}	\s^{\m}_{\a	\dot{\b}}
\overline{c}^{\dot{\b}}
\ee
\be
\delta	{\tilde	A} 	=
c^{\a}	{\tilde	\y}_{	\a}
-	\ov{
F} +	2 m	 A
+ 3  g 	A A
\ee
\be
\delta	{\tilde	\y}_{	\a}	=
	\pa_{\m} {\tilde A} \sigma^{\m}_{\;\a	\dot	\b}
\ov{c}^{\dot	\b}   +	{\tilde	F} 	c_{\a}
+\sigma^{\m}_{\;\a	\dot	\b}	\pa_{\mu}
\ov{\y}^{\dot	\b}
 + 2 m
 	\y^{\a }
+ 6  A   \y_{\a}
\ee
\be
\delta	{\tilde
F}	= \pa_{\m}
{\tilde	\y}^{	\a}	\s^{\m}_{\a	\dot{\b}}
\overline{c}^{\dot{\b}}
+ \Box	{\ov	A}   + 2 m	 F + 6 g 	 (	A 	F  +
\y^{\a	 }	\y_{\a} 	)
\ee
\be
\d \f_{\a} = {\ov c}^{\dot \b} W_{\a \dot \b}
\ee
 \be
\d W_{\a \dot \b } =\pa_{\m} \f_{\a}
{\ov \s}^{\m}_{\dot \b \g} c^{\g}
+ \c_{\a} {\ov c}_{\dot \b}
\ee
 \be
\d \c_{\a} = \pa_{\m} W_{\a}^{ \;\; \dot \b}
{\ov \s}^{\m}_{\dot \b \g} c^{\g}
\la{mew2}
\ee
Note that we have kept a simple non-linear term in
the equation of motion, so that we can see its effect.

\section{General Form of $P_{\F}(G=1,D=2) $ }

Now let us write down the most general Lorentz
invariant integrated local polynomial, linear in the $\F$
superfield, of ghost charge 1 with the very low dimension 2.
It is:
\[
P_{\F}(G=1,D=2) = \int d^4 x
\Bigl \{ \c^{\a} c_{\a} [ g_1 {\ov A} + g_2  A ]
\]
\[
+  W^{\a \dot \b}  [  g_3  c_{\a} {\ov \y}_{\dot \b}
 + g_4     \y_{\a} {\ov c}_{\dot \b} ]
+  W^{\a \dot \b}     c_{\a} {\ov c}_{\dot \b} [ g_5 {\tilde A}
 +  g_6 {\tilde {\ov A} } ]
\]
\[
+  \f^{\a}   c_{\a} [ g_7 {\ov F}
 + g_8    F+ g_9 m A +g_{10}  m {\ov A}
 + g_{11}    A^2 + g_{12} A {\ov A}     + g_{13}  {\ov A}^2 ]
\]
\be
 +  \f^{\a}   c_{\a} [ g_{14} {\tilde \y}^{\b}   c_{\b}
+  g_{15}  {\tilde {\ov \y}}^{\dot \b}   {\ov c}_{\dot \b} ]
+
\f^{\a}     \s^{\m}_{\a \dot \b } {\ov c}^{\dot \b }
[ g_{16}  \pa_{\m} A  + g_{17}  \pa_{\m} {\ov A } ]
\ee
We note that:
\begin{enumerate}
\item
The dimension of the source superfield $\tilde S$ is
two whereas the dimension of the superfield $S$ is
only one.  This limits the possibilities of constructing
terms with more than one $c_{\a}$, since such terms
have to be accompanied by $\tilde S$ fields to
compensate the higher ghost charge associated with
more than one $c$.
\item
There are more terms with $\f$ than there are with
$W$ and more terms with $W$ than there are with
$\c$.  This is due to the fact that $\f,W$ and $\c$
occur linearly.  The number of  possibilities is greater for
the terms accompanying $\f$ than for the
terms accompanying $\c$  because $\f$ has lower dimension
than $\c$.
\item
This general polynomial will not arise in perturbation
theory--only the subspace $K_{\F}(G=1,D=2) \subset
P_{\F}(G=1,D=2)$  of such
polynomials which satisfy
\be
\d K_{\F}(G=1,D=2) =0
\la{exact}
\ee
could be expected to arise.  One could simply solve this
set of equations for the maximal free set of coefficients
$g_i$ , and one would call the resulting polynomial
$K_{\F}(G=1,D=2)$  the set of all $\d$-closed
polynomials. But there is another way to  construct this
space of course, which  is the  whole point of cohomology
theory.  We will do that below.
\item
The number of terms here will rapidly increase as the
dimension D in $P_{\F}(G=1,D=2)$ rises.  We need to
be able to pick out some special minimum number of  terms
to calculate the physically relevant part while ignoring the
physically irrelevant boundary (image of $\d$) terms.
\end{enumerate}

\section{Counterterms and the Anomaly}
\la{countert}
Continuing with our dimension 2 example, we now
write down the most general polynomial linear in $\F$
with ghost charge 0 and dimension 2:
\be
P_{\F}(G=0,D=2) = \int d^4 x
\{
    e_1 \f^{\a} \y_{\a} + e_2 \f^{\a} c_{\a} {\tilde A}
+ e_3  \f^{\a} c_{\a} {\tilde{\ov A} }
\}
\la{actionsm}
  \ee
Now we can use our cohomology result, which
states that the most general solution
of the equation (\ref{exact}) takes the form:
\be
K_{\F}(G=1,D=2)
 = \d P_{\F}(G=0,D=2) + e_4 H_{\F}(G=1,D=2)
\ee
and we recall that in general we could expect to
get the following result in perturbation theory
\be
 \d \G_{\F} =K_{\F}(G=1,D=2).
\ee
For the present case, we get from (\ref{actionsm})
and the above transformations in (\ref{mew1})--
(\ref{mew2}):
\[
K_{\F}(G=1,D=2)
=
\int d^4 x
\{
    e_1
[
{\ov c}_{\dot \b} W^{\a \dot \b}
\y_{\a} + \f^{\a}
(  \pa_{\m}	  A  \s^{\m}_{\a	\dot{\b}}
\overline{c}^{\dot{\b}} +	F	c_{\a}  )
\]
\[
+ e_2
[
 {\ov c}_{\dot \b} W^{\a \dot \b}
 c_{\a} {\tilde A}
+ \f^{\a} c_{\a}
(c^{\b}	{\tilde	\y}_{	\b}
-	\ov{
F} +	2 m	 A
+ 3  g 	A^2 )
]
\]
\[
+ e_3
[
{\ov c}_{\dot \b} W^{\a \dot \b}
 c_{\a} {\tilde{\ov A} }
+ \f^{\a} c_{\a}
( {\ov 	c}^{\dot \a}
{\tilde {\ov \y}}_{\dot \a}
- F + 2 m {\ov A} + 3 g {\ov A}^2
)]
\]
\be
+ e_4 [ \c^{\a} c_{\a} {\ov A} +
W^{\a \dot \b } c_{\a} {\ov \y}_{\dot \b} +
\f^{\a} c_{\a} {\ov F} ]
\}
\ee
In this particular case, the term
$\c^{\a} c_{\a} {\ov A}$ occurs with the
coefficient $e_4$, which is the physically
meaningful coefficient of the
anomaly.  What term in $\G$ could give rise to
this term?  The only way that this could arise in
the expression  $\d \G_{\F}$ is from the terms
\[
e_4 \int d^4 x \c^{\a} c_{\a} {\ov A}
=
	\int d^4 x
[\pa_{\m} {\ov A} \sigma^{\m}_{\;\a	\dot	\b}
 {c}^{ 	\a}   \fr{\d}{ \d  \ov{\y}_{\dot	\b}  } ]
 \G(\c, {\ov \y} )
\]
\be
+	\int d^4 x
[\Box {\ov A} \fr{\d}{ \d  \tilde{F} } ]
 \G(\c,c , {\tilde  F} )
\la{basic}
\ee
 The derivative in the functional deriviative operator
could conceivably convert the non-local
functionals  $\G(\c, {\ov \y})$ and $ \G(\c,c , {\tilde  F} )$
into the
local cohomologically nontrivial term
$ \int d^4 x \c^{\a}
c_{\a} {\ov A} $  with a
non-zero coefficient $e_4$.

 The  coefficient $e_4$ could also be obtained as
the coefficient of the term $W^{\a \dot \b } c_{\a} {\ov
\y}_{\dot \b}$ according to
\[
e_4 \int d^4 x W^{\a \dot \b } c_{\a} {\ov
\y}_{\dot \b} =
	\int d^4 x
[\pa_{\m} {\ov \y}^{\dot \b}  \sigma^{\m}_{\;\a	\dot	\b}
 {c}^{ 	\a}    \fr{\d}{ \d  \ov{F } }]
 \G(W, {\ov F})
\]
\[
+\int d^4 x
[\pa_{\m} {W}^{\a \dot \b}
{\ov \sigma}^{\m}_{\;\dot \b	 \b}
 {c}^{ 	\b}    \fr{\d}{ \d  \c^{\a} } ]
 \G(\c, {\ov \y} )
\]
\be
+\int d^4 x
[\pa_{\m} {\ov \y}^{ \dot \b}
{\ov \sigma}^{\m}_{\;\dot \b	 \b}
\fr{\d}{ \d  {\tilde \y}_{\b} } ]
 \G(W,c, {\tilde \y} )
\la{notbasic}
\ee
This computation is harder than the one involving
$\c$ because it requires knowledge of three parts
of $\G$ rather than two.  It is even harder to
calculate the coefficient  $e_4$
using the term  $\f^{\a} c_{\a} {\ov F}$, because
it gets contributions  from the boundary term with
coefficient  $e_2$.   To obtain  $e_4$ from
$\f^{\a} c_{\a} {\ov F}$ one
would need to calculate the coefficient  $e_2$
of the boundary
terms also and then subtract.  So clearly the $\c_{\a}$
method is the easiest. Let us summarize some of the
lessons
we have found  here:
\begin{enumerate}
\item
One could expect that the coefficient of
the anomaly is generated most easily from
calculation of the $\c_{\a}$ terms in the action,
since there are fewer of them and they are more
likely to be directly linked to the anomaly coefficient.
In higher dimensional cases one would calculate
at a minimum the terms
 $\G(\c, {\ov A}, {\ov A}, \cdots, \ov{\y} )$
and
 $\G(\c,c, {\ov A}, {\ov A}, \cdots, {\tilde F} )$
with one $\c$ and one $\ov {\y}$ or ${\tilde F}$
and all available  numbers of $\ov A$ fields--see
section (\ref{masssec}).
\item
A detailed look at the way the boundary terms
contribute to the coefficients will always be
needed for specific cases, and this will become
increasingly complicated as the dimension increases.
\item
The essential equation for finding the coefficient of
the anomaly is (\ref{basic}) and its generalization to
higher dimensions and more fields.
\item
Attempting to find the coefficient of the anomaly
using equations like (\ref{notbasic}) involves more
work and is more likely to get mixed up with
boundary terms as the dimension increases.
Of course, it would be desirable to confirm that
this gives the same result if one first finds a non-zero
result for the form (\ref{basic}).
\item
The present example is useful because it is easy to
generate complete expressions which
give results of general validity, but it has too low
a  dimension to have any application to
actual Feynman diagrams, since there is no
composite antichiral spinor superfield with
dimension $\fr{1}{2}$, which is what the present
case would require.  Indeed the lowest dimension
 composite antichiral spinor superfield is
(\ref{e1}) which has dimension $3\fr{1}{2}$,
and  even it still looks too low in dimension
(from the point of view of divergence of the relevant
Feynman diagrams) to have a reasonable chance of
developing an anomaly.
\item
The discussions in sections (\ref{masssec}) and (\ref{secexample})
are now seen
to deal with the important part of the calculation
for practical purposes--namely the  $\c_{\a}$ part.
\end{enumerate}

\section{Conclusion}

Our result is that when one considers the complete BRS
operator for gauge-fixed
supersymmetric Yang-Mills theory coupled to chiral matter
in four spacetime dimensions with spontaneous breaking of
gauge symmetry, but not of supersymmetry, one finds that
there is a subspace $H_{\rm Special}$
of the cohomology space which was described in section
(\ref{hspecial}) and (\ref{suppot}) ,  and which can be dealt with
along the lines of section (\ref{masssec}) and (\ref{countert}).

Supersymmetry anomalies of the simplest kind envisioned
here
would involve superfields which are massless after
spontaneous
symmetry breaking but which do not contain
Goldstone  bosons.  A perfect example of such a superfield
is the neutrino superfield in the standard model.
See \ci{tamu47} for more details of the current
analysis in the context of the standard model.
For reasons explained above, that kind of anomaly can only
occur if the gauge symmetry is spontaneously broken.

If this were the end of the story, it would be interesting
but also rather worrying from the phenomenological point of
view, because it is hard to imagine that all supersymmetry
breaking arises from an anomalous mixing of massive
superparticles with the components of the neutrino superfield.
Fortunately, there are a large number of
superanomalies involving massive fields also, but
these are a little more involved to separate from
non-physical parts which vanish by the equations of motion.
For these other cases, it is not so obvious that
one needs spontaneous gauge symmetry breaking to get the
anomalies, but  again there is probably  a need to `mix up'
fields at a gauge-chiral vertex that seems most easily
accomplished by a spontaneously broken theory.  This
needs further investigation.

One might be concerned that the equation of motion of the
superfield sources $\F_{\a}$ have not been used in the above.
Actually these fields have some complicated problems of their
own, but it does not appear that including these complications
would be likely to make the cohomology space empty or to
significantly change the present results.  It should also
be mentioned that a non-trivial transformation of the
$\F_{\a}$ sources under $U(1)$ is probably possible,
which might be needed to explain the supersymmetry breaking of
the charged particles.

In conclusion, it still seems to be possible that these results
are the start of a phenomenologically interesting explanation
of the breaking of supersymmetry for all observed particles. If
this is indeed the explanation, then there must be a wealth
of predictions  available from pertubation theory
with a minimum of uncalculable non-perturbative
dynamical assumptions.  The first step to
test this possibility is to calculate some of the
simpler examples to see whether these anomalies do indeed appear
with non-zero coefficients.  The main issue there seems to
be the question of an appropriate regularization to use
for this problem.

Acknowledgments:  I would like to thank my collaborators Ruben
Minasian and Joachim Rahmfeld for many useful ideas, Ramzi
Khuri for some very useful discussions about integrals,
Heath Pois for many helpful remarks about
phenomenological models of supersymmetry, Chris Pope
for  helping us to recognize the representations
that arose in  the
solution of the higher spin problems and Mike Duff for
his continued insistence that a supersymmetry anomaly be
computed.

\end{document}